\documentclass[twocolumn,showpacs,amsmath,amssymb,prc,superscriptaddress,nofootinbib]{revtex4-1}

\usepackage[pdftex]{graphicx}
\usepackage{epstopdf}

\usepackage{dcolumn}
\usepackage{bm}
\usepackage{times}
\usepackage{siunitx}
\usepackage{booktabs}
\usepackage{etoolbox}
\usepackage{braket}
\usepackage{threeparttable}
\usepackage{rotating}
\usepackage{multirow}
\usepackage{subcaption}
\usepackage{aas_macros}
\usepackage{dirtytalk}
\robustify\itshape
\allowdisplaybreaks

\AtBeginDocument{
\heavyrulewidth=.08em
\lightrulewidth=.05em
\cmidrulewidth=.03em
\belowrulesep=.65ex
\belowbottomsep=0pt
\aboverulesep=.4ex
\abovetopsep=0pt
\cmidrulesep=\doublerulesep
\cmidrulekern=.5em
\defaultaddspace=.5em
}

\newcolumntype{.}{D{.}{.}{1}}
\newcolumntype{X}{D{X}{X}{1}}

\newcommand{\printthis}[2][true]{%
\ifbool{#1}{%
#2}{}}

\newcommand{\aomp}[0]{$\alpha$-OMP }

\newcommand{\ra}[1]{\renewcommand{\arraystretch}{#1}}

\newcommand{\reac}[4]{#1$($#2$,$#3$)$#4 }
\newcommand{\iso}[2]{$^{#1}$#2}
\newcommand{\Sraa}{\reac{\iso{86}{Sr}}{$\alpha$}{$\alpha$}{\iso{86}{Sr}}}
\newcommand{\Sran}{\reac{\iso{86}{Sr}}{$\alpha$}{$n$}{\iso{89}{Zr}}}
\newcommand{\Srap}{\reac{\iso{86}{Sr}}{$\alpha$}{$p$}{\iso{89}{Y}}}
\newcommand{\Srag}{\reac{\iso{86}{Sr}}{$\alpha$}{$\gamma$}{\iso{90}{Zr}}}

\begin{document}
\def\sun{\odot}

\title{Bayesian Analysis of the $^{86}$Sr$(\alpha, \alpha)$ Reaction to Constrain the $^{86}$Sr$(\alpha, n)$ Cross Section at Astrophysical Energies}

\author{C.~Marshall}
 \affiliation{Department of Physics and Astronomy, University of North Carolina at Chapel Hill, Chapel Hill, NC 27599, USA}
\affiliation{Triangle Universities Nuclear Laboratory, Durham, NC
  27708, USA}

\author{T.~Lansing}
\altaffiliation[Present Address: ]{United States Navy, Charleston, SC 29404, USA}
\affiliation{The College of Wooster, Wooster, OH 44691, USA}

\author{D.~Gribble}
 \affiliation{Department of Physics and Astronomy, University of North Carolina at Chapel Hill, Chapel Hill, NC 27599, USA}
\affiliation{Triangle Universities Nuclear Laboratory, Durham, NC
  27708, USA}

\author{R.~Longland} \affiliation{Department of Physics, North
  Carolina State University, Raleigh, NC 27695, USA}
\affiliation{Triangle Universities Nuclear Laboratory, Durham, NC
  27708, USA}

\author{A.~Psaltis}
\affiliation{Department of Physics, Duke University, Durham, NC 27708, USA}
\affiliation{Triangle Universities Nuclear Laboratory, Durham, NC
  27708, USA}

 \author{K.~Setoodehnia}
 \affiliation{Department of Physics, Duke University, Durham, NC 27708, USA}
\affiliation{Triangle Universities Nuclear Laboratory, Durham, NC
  27708, USA}

\begin{abstract}
The Alpha Optical Model Potential (\aomp \!) is a phenomenological approach used to describe elastic scattering where multiple reaction channels are open. It is one of the most critical inputs for the calculation of thermonuclear reaction rates in explosive stellar environments, but uncertainties within the $\alpha$-OMP lead to imprecise predictions hindering comparisons between calculations and observations. In order to improve the precision of the $\alpha$-OMP, additional nuclear physics data are required. In this paper, a measurement of the $^{86}$Sr($\alpha$, $\alpha$) elastic scattering cross section at multiple energies is reported. A local optical potential is constructed via a fully Bayesian analysis of the elastic scattering data. The resulting uncertainties on the low energy cross sections relevant to nuclear astrophysics are then calculated and shown to be on the order of $50 \%$.

\end{abstract}

\maketitle


\section{Introduction}
Observations of elemental abundances in metal-poor stars located in the Galactic halo show significant scatter for elements of the first rapid neutron capture process peak (r-process), between strontium (Sr, $Z=39$) and silver (Ag, $Z=47$)~\cite{Frebel2018}, but follow a robust r-process pattern for heavier elements~\cite{Horowitz2019}. Currently, the astrophysical origin of these observations is unclear. One possible scenario is that these intermediate mass elements are the product of nucleosynthesis occurring in the ejecta of neutrino-driven ($\nu$-driven) winds from core collapse supernovae (CCSN)~\cite{Psaltis2024}. In the case of neutron rich ejecta, temperatures of 2-5 GK following an $\alpha$-rich freezeout from the nuclear statistical equilibrium (NSE) would drive nucleosynthesis in the A = 60 - 110 region primarily through $(\alpha, n)$ reactions involving nuclei a few units away from stability. This scenario is known as the weak r-process or $\alpha$-process~\cite{Woosley1992, Hoffman1997, Arcones2011}.

For medium-mass nuclei at temperatures above a few gigakelvin, $(\alpha, n)$ reactions { will proceed through a large number of states in the compound nucleus} and, as a result, are well described by the predictions of the Hauser-Feshbach statistical model \footnotemark[1] \cite{hauser_feshbach_1952}. Above the neutron threshold, neutron emission is highly favored when compared to other open reaction channels, typically making the cross section of $(\alpha, n)$ reactions at least an order of magnitude larger than those of $(\alpha, p)$ or $(\alpha, \gamma)$ reactions. As a result, the total reaction cross section and the $(\alpha, n)$ cross section are nearly equal, and the former is sensitive only to the Alpha Optical Model Potential (\aomp\!), making it the key input in Hauser-Feshbach calculations \cite{Mohr-2016}. However, these $(\alpha, n)$ reactions still occur below the Coulomb barrier, where it has been found that the many available \aomp produce cross sections that can vary by an order of magnitude \cite{Mohr-2016,Pereira-2016}. These cross section uncertainties propagate to the corresponding thermonuclear reaction rates, and, hence, to the abundance predictions of CCSN models, stymieing efforts to connect observations of metal-poor stars to specific astrophysical conditions \cite{Bliss_2020, Psaltis2022}.

Save for a few exceptions, most optical potentials used in the literature are phenomenological in nature, where parameterized real and imaginary potentials are adjusted to reproduce elastic scattering angular distributions, total reaction cross sections, and analyzing powers \cite{hodgson_1994}. It is critical to note that the sensitivity studies of Refs.~\cite{Bliss_2020, Psaltis2022} account only for uncertainties inferred from comparing the predictions of a handful of \aomp with one another, and not for any parametric uncertainty arising from the \textit{phenomenologically} derived potential parameters. An unfortunate situation arises from the lack of uncertainty quantification for these \aomp\!: the predictions of individual optical models do not include uncertainty estimates, meaning one cannot assess their precision when predicting $(\alpha, n)$ cross sections, nor can one determine the degree to which different models actually disagree.

\footnotetext[1]{Throughout this paper it will be necessary to use ``Hauser-Feshbach'' to refer to the statistical theory of nuclear reactions in order to distinguish it from the Bayesian statistical model that is the primary focus of the paper.}

Recent efforts by the nuclear astrophysics community have focused on addressing the scarcity of relevant $(\alpha, n)$ cross section data, particularly for key isotopes and neighboring nuclei. These experiments have employed both stable and radioactive ion beams with a variety of techniques, including activation methods~\cite{Kiss2021, Szegedi2021}, active-target systems~\cite{Ong2022, Fougeres2024, Fougeres2025}, and recoil mass separators~\cite{Williams2025}. However, in these studies the measured $(\alpha, n)$ cross sections are used only to discriminate between available \aomp, not to improve or determine new \aomp parameters. 

Several previous studies have taken direct steps towards determining \aomp parameters by measuring elastic scattering data from a single or few isotopes, with mixed success. For example, the high precision scattering data of Ref.~\cite{mohr_1997} for $^{144}$Sm$(\alpha, \alpha)$ failed to predict the $^{144}$Sm$(\alpha, \gamma)$ cross sections measured in Ref.~\cite{Somorjai-1998}. A similar result was found for $^{94}$Mo$(\alpha,\alpha)$ and $^{94}$Mo$(\alpha,n)$ in Ref.~\cite{fulop-2001} and Ref.~\cite{Rapp_2008}, respectively. The larger scale study of Ref.~\cite{Palumbo_2012} found some improved predictions with their potential, but just as many worsened predictions for $\alpha$-induced reactions on $^{106}$Cd, $^{118}$Sn, and $^{120, 124, 126, 128, 130}$Te. Again, absent from these studies is a robust determination of the uncertainty that provides confidence intervals for the observed agreement or disagreement. 

In this paper, we report on a measurement of \Sraa that has been analyzed using a first-of-its-kind Bayesian statistical model. Our method allows us to extract \aomp parameters and examine their influence on the \Sran cross section taking into account all parameter correlations. These correlations are in turn propagated through Hauser-Feshbach calculations to show the extent to which a single, modestly precise scattering experiment can constraint the \aomp and how cross section measurements can be combined with scattering data to improve the predictive power of the model. In doing so, we aim to illustrate the limitations of elastic scattering data and outline possible directions for future experiments.

The paper is organized as follows: Section \ref{sec:experiment-details} details our experimental setup for the \Sraa measurement, Section \ref{sec:analysis} introduces all relevant details for our analysis including a detailed description of the Bayesian model, Section \ref{sec:results} presents and gives context to our results, and Section \ref{sec:conclusions} summarizes this work with suggestions for future work.

\section{Experiment Details}
\label{sec:experiment-details}

The $^{86}$Sr$(\alpha, \alpha)$ experiment, as well as its ancillary measurements, was carried out at Triangle Universities Nuclear Laboratory (TUNL) using the 10 MV TUNL FN tandem accelerator. The TUNL tandem facility has an analyzing magnet that serves several beamlines. Our experiments were all carried out using the $52^{\circ}$ line and its associated scattering chamber. The scattering chamber is approximately 60 cm in diameter and has two independent rotating plates on the top and bottom of the chamber. These plates can be fully rotated without breaking vacuum and can accommodate up to 10 total detectors (5 top and 5 bottom). For these experiments three silicon surface barrier detectors (SSB) were used to measure angular distributions. All of the detectors had an active area of 150 mm$^2$ and a depletion depth of at least 300 $\mu$m in order to ensure that $21$-MeV $\alpha$-particles would be stopped within their depletion regions. One rectangular collimator was used for each detector as a balance between maximizing their solid angles, reducing background induced from scattered beam, and minimizing the sensitivity to beam and detector misalignments. The geometric solid angle of each detector was found to be ${0.98(8)}\,$msr. Detector signals were transmitted through a Mesytec MSI-8 preamplifier and shaper \cite{Mesytec}. The output was sent into a CAEN V785 peak sensing analog-to-digital converter (ADC) \cite{Ortec_ADC} that was triggered off the common timing output of the MSI-8. Electronic dead time was measured with a pulser and was below 10\% for all the data used in this study. Data was recorded for offline analysis using the MIDAS data acquisition software \cite{midas}.  

\subsection{Magnet Calibration}
\label{sec:magnet-cal}
The analyzing magnet was energy calibrated for the $52^{\circ}$ beamline using the $E_p\!=\!1748$-keV resonance in the $^{13}$C$(p, \gamma)$ reaction \cite{Marion_1966, Ajzenberg_1991}.
A yield curve was measured with a large volume NaI detector placed on top of the scattering chamber. Average proton currents were on the order of $400$ pnA. A self-supporting, 40 $\mu$g/cm$^2$ $^{\text{nat}}$C carbon foil target was used. The thick target yield curve was used to determine the magnetic constant using the relativistic equation:
\begin{equation}
    B \rho = \frac{1}{q} \sqrt{\frac{m}{c_{mag}}\bigg[\frac{E^2}{2mc^2u} + E\bigg]},
\end{equation}

where $B$ is the measured field of the NMR probe of the analyzing magnet in mT, $\rho$ is the bending radius for the $52^{\circ}$ line in meters, $m$ is the mass of the beam in amu, $E$ is the beam energy in keV, $q$ is the charge state of the beam, and $u$ is amu in units of keV$/c^2$ = $931.494 \times 10^3$ keV$/c^2$. We determined $c_{mag} = 0.02404(12)$ $(\text{keV} \cdot u)/ (\text{m}^2 \cdot \text{mT}^2)$, where the uncertainty is statistical only. Due to the low energy at which this calibration was performed and our neglect of hysteresis, we estimate an additional uncertainty at each of our beam energies based on the reproducibility of the yield and comparison to previous energy calibrations of the same beamline. The beam energies for the \Sraa experiment where determined to be $12077 \pm 6(stat.) \pm 40(syst.)$, $18101 \pm 9(stat.) \pm 60(syst.)$, and $20530 \pm 10(stat.) \pm 70(syst.)$ keV.   

\subsection{Target Fabrication}
\label{sec:target-fab}
SrCO$_3$ powder material enriched to $96.89\%$ in $^{86}$Sr was deposited via thermal evaporation onto self-supporting carbon foils of natural isotopic abundance. The carbon foils were acquired from the Arizona carbon foil company \cite{acfmetals} and were quoted as having a thickness of $40(4) $ $\mu$g/cm$^2$. Rutherford back scattering (RBS) was carried out during a separate beam time to determine target properties. A $^4$He$^{++}$ beam was accelerated to $2$ MeV and backscattered $\alpha$-particles were detected at $165^{\circ}$ in the lab frame. Examining both Oxygen and Strontium elastic scattering peaks indicated an oxide layer was produced during the evaporation with a Sr:O ratio of $\approx$1:1. An areal density for the SrO layer of $24.6(25)$ $\mu$g/cm$^2$ was deduced from the areas of these peaks. A small amount of tantalum was found to be present in the target during the main experiment, consistent with a small amount of the boat material being deposited during the evaporation of the SrCO$_3$ material. The amount was too small to be reliably observed during the RBS experiment.

\subsection{$^{86}$Sr$(\alpha, \alpha)$}
\label{sec:86Sr-alpha-alpha}
Beams of $^{4}$He$^{++}$ were accelerated to energies of approximately $12$, $18$, and $21$ MeV (for precise energies see Sec.~\ref{sec:magnet-cal}). Beam currents on target during the experiment were typically around $45$ pnA. Currents were measured with a Faraday cup located approximately 1 m downstream from the target and suppressed with a voltage of -300 V. After the experiment, the absolute current readings were found to be unreliable. It was determined that the suppression voltage had been switched inadvertently resulting in an incorrect current reading. The detector on the bottom plate of the chamber was left stationary at each energy to monitor target degradation, with $\theta_{Lab} = 150^{\circ},70^{\circ}, 80^{\circ}$ at $12$, $18$, and $21$ MeV respectively. No sign of degradation was seen over the course of the experiment. The two detectors on the top plate were rotated to measure angular distributions from $30^{\circ}$-$150^{\circ}$ in $5^{\circ}$ increments. The target ladder was rotated by $45^{\circ}$ relative to the beam for detector angles between $60^{\circ}$ and $120^{\circ}$, where the target ladder would shadow the solid angle. 

\begin{figure*}
    \centering
    \includegraphics[width=0.8\linewidth]{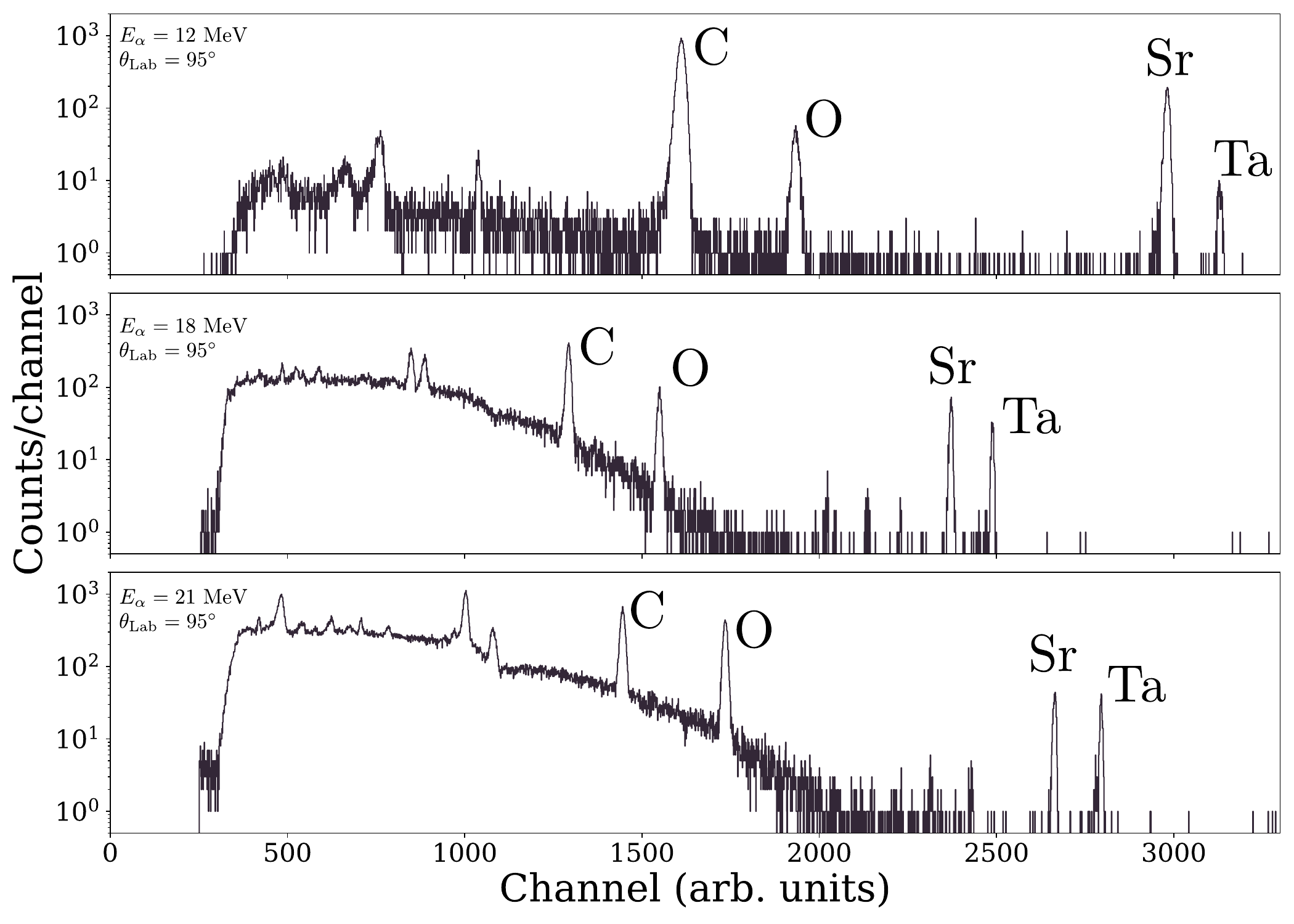}
    \caption{Silicon detector spectrum at $\theta_{\text{Lab}} = 95^{\circ}$ for 12 (top), 18 (middle), and 21 (bottom) MeV. Tantalum
    contamination came from the boat used to evaporate the $^{86}$SrCO$_3$ powder. Note that the shaping amplifier gain was adjusted after the
    12 MeV spectrum but was then held constant. The peaks at lower $\alpha$-particle energy were not identified and are irrelevant to the current study.}
    \label{fig:12mev-spectrum}
\end{figure*}

\section{Analysis}
\label{sec:analysis}

$\alpha$-particle spectra from \Sraa were constructed for each angle and energy using the software package \texttt{sauce} \cite{sauce_1}. Typical spectra from the silicon detectors are provided in Fig.~\ref{fig:12mev-spectrum}. The elastic scattering peaks from $^{86}$Sr were clearly separated from contaminates at all angles and readily identifiable. No attempt was made to energy calibrate the spectra or to determine the origin of peaks that did not arise from elastics scattering. Peak areas were determined by fitting a Gaussian peak shape with a linear or quadratic background using the \texttt{fityk} package \cite{Wojdyr_2010}. As mentioned in Section~\ref{sec:86Sr-alpha-alpha}, we could not determine the absolute number of beam particles incident due to an unreliable charge integration. Nevertheless, we constructed angular distributions using the measured charge accumulation and target thickness (Sec.~\ref{sec:target-fab}), but left the absolute scaling as a free parameter in our optical model calculations. 

It is our goal to compare, in as complete a fashion as possible, the \Sran cross sections predicted base on our \Sraa data with those of many optical models. We have carried out Bayesian inference on the \Sraa data; however, it is computationally intensive to propagate the resulting parameter uncertainties through Hauser-Feshbach calculations at many energies. Instead, we choose to access the impact of elastic scattering data on the predicted cross section by focusing on a \textit{single} energy for the \Sran reaction ($E_{\alpha} = 6.2$ MeV). By focusing on a single quantitative measure of variation caused by the \aomp\!, we sidestep the larger issue of whether the elastic scattering data are able to \textit{accurately} predict the low energy cross section. Comparisons to and impacts of the measured $^{86}\text{Sr}(\alpha, n)$ data will be discussed in Sec.~\ref{sec:potential-impacts-cross}. We are also assuming that the variation at this single data point will be indicative of the overall reaction rate uncertainty. In the subsections that follow, we will detail each of the calculations necessary to analyze and interpret our experimental data.



\subsection{Parameterization of the \aomp}
\label{sec:aomp-param}

It is insufficient to analyze the \Sraa angular distributions without first justifying our choice of parametrization of the \aomp. { It has been frequently claimed in the past that elastic scattering data fails to constrain the low energy reaction cross sections \cite{mohr-2021}; however, elastic scattering studies often adopt various \aomp formalisms \cite{Somorjai-1998, fulop-2001}, introducing additional systematic uncertainties that make quantitative comparison difficult. In this work, we focus on the effects of a single optical model formalism to investigate parametric uncertainties.}

The \aomp adopted for this study is a simple 6 parameter version similar to those employed in Ref.~\cite{McFadden-1966, avrigeanu-1994, avrigeanu-2014, Nolte_1987}. It is defined as:
\begin{equation}    
\label{eq:omp-this-study}
\mathcal{U}(r, E) = V_c(r; r_c)-V(E)f(r; r_0, a_0) -iW(E)f(r; r_i, a_i) \\,
\end{equation}
where $V_c(r; r_c)$ is the Coulomb potential of a uniformly charged sphere and $f(r)$ is the Woods-Saxon form factor:
\begin{equation}
  \label{eq:ws_pot}
  f(r; r_0, a_0) = \frac{1}{1 + \exp\big(\frac{r-r_0A_t^{1/3}}{a_0}\big)}.
\end{equation}
The form factor has two parameters: $r_0$ is the radius and $a_0$ is the diffuseness. $A_t$ is the mass number of the target. We have included only a volume imaginary potential due to historical precedence \cite{McFadden-1966, Nolte_1987, avrigeanu-1994} and the limited amount of data under consideration for this experiment. We assume the Coulomb radius to have the common value of $r_c=1.3$ fm.

Since our elastic scattering measurements were carried out well above the energies of interest to astrophysics, our model would be incomplete without specifying the energy dependence of the potentials. We only allow the real and imaginary depths to be energy dependent, as was done in Refs.~\cite{Nolte_1987,avrigeanu-1994,avrigeanu-2014,Demetriou_2002}. For the real potential we adopt a linear dependence:
\begin{equation}
  \label{eq:real-depth-linear}
    V(E) = V_0 - V_1E.
\end{equation}
For the imaginary volume term two parameterizations are used. Each parameterization is treated as a separate model (i.e. a separate fitting process) in the analysis. First is a linear form identical, except for a sign, to the real potential:
\begin{equation}
  \label{eq:imaginary-depth-linear}
    W(E) = W_0 + W_1E, 
\end{equation}
and the second is the Fermi form:
\begin{equation}
  \label{eq:imaginary-depth-fermi}
  W(E) = W_0 \bigg(1 + \exp{\frac{W_1 - E}{W_2}} \bigg)^{-1}
\end{equation}
Both of these forms follow other \aomp studies \cite{Le-2022, Somorjai-1998, fulop-2001}. Furthermore, the Fermi form of energy dependence is not unique to the \aomp and has been adopted in the $p,n$ global optical model of Ref.~\cite{varner-1991} and $^3$He$,t$ global optical model of Ref.~\cite{Liang_2009}.

\subsection{Hauser-Feshbach Model Calculations}
\label{sec:stat-model}

All Hauser-Feshbach model calculations of the \Sran cross section were performed with the TALYS 2.0 code \cite{koning-2023}. With the exception of the \aomp and level densities, the default TALYS' parameters were used. For the level density, the back shifted Fermi gas model (BSFG) was selected due to the observations of Ref.~\cite{hamad-2022}. In that work a closer agreement was found between existing experimental data and the predictions of {Hauser-Feshbach} model calculations when using the BSFG level density instead of the TALYS default. We observed the same trend for $^{86}$Sr$(\alpha, n)$ when comparing to the data of Ref.~\cite{levkovski_1991}. TALYS includes 8 \aomp by default.

The reaction rate will be most sensitive to the cross section at energies where its integrand is at a maximum:
\begin{equation}
\label{eq:rr-integrand}
    \langle \sigma v \rangle = \bigg(\frac{8}{\pi \mu}\bigg)^{1/2} \frac{1}{(kT)^{3/2}} \int_0^{\infty} E \sigma(E) e^{-E/kT} dE,
\end{equation}
where $k$ is Boeltzmann's constant, $E$ is the center-of-mass energy, $\mu$ is the reduced mass of the reacting particles, and $T$ is the temperature of the plasma. 
It must be stressed that for the $^{86}$Sr$(\alpha, n)$ reaction at temperatures of $2$ GK the concept of a ``Gamow Window" is not applicable due to the cross section no longer being well described by $s$-wave barrier penetrability \cite{newton-2007, Rauscher-2010}. Using TALYS with the parameters described above along with the McFadden and Satchler Optical Model, we find Eq.~\ref{eq:rr-integrand} to reach a maximum around 5.9 MeV in the center-of-mass (compared to the 5.4 MeV of the Gamow peak).
This corresponds to a laboratory energy of $E_{\alpha} = 6.2$ MeV, which will be used for all quantification of the cross section uncertainty and as a stand-in for the reaction rate uncertainty. { At 6.2 MeV, two other reaction channels are open in addition to \Sran ($Q=-5.292$ MeV): \Srag ($Q=6.674$ MeV) and \Srap ($Q=-1.676$ MeV).}

Table~\ref{tab:talys-model-calcs} provides the predictions of the 8 \aomp that are included in TALYS, displaying a variation of two orders of magnitude. For comparison, we have also included the predictions of the Atomki-V2 potential calculated using a modified version of TALYS 1.8 \cite{mohr-2021}. Table~\ref{tab:talys-model-params} lists the optical model parameters that produced the statistical model calculations. Qualitatively it is difficult to determine exactly why each of these models produces such wildly different predictions. Table~\ref{tab:talys-model-calcs} also shows what type of energy dependence is used for the imaginary potential. All models used in TALYS, with the exception of the McFadden and Satchler parameters, either assume a linear or Fermi type energy dependence, further justifying our decision to consider only those two forms. { We also note that the choice of the BSFG level density has a reasonably large impact on these predictions. Compared to the other models in TALYS, the BSFG model produces an \Sran cross section a factor of 1.4 to 1.5 higher at 6.2 MeV and also predicts a \Srap branching ratio a factor of 30 weaker. Since this shift in cross section is nearly constant among optical models, we do not consider this uncertainty further in this study.}

\begin{table*}
\ra{1.3}
\begin{tabular}{@{}llll@{}}
\toprule\toprule
\aomp& TALYS ID  &$\sigma_{(\alpha, n)}$ ($\mu$b) & Comment \\ \hline
 \citet{koning-2023}         &1         &1.002 & \\
 \citet{McFadden-1966}       &2         &0.642 & Not a global model, $W$ only.\\
 \citet{Demetriou_2002}      &3         &0.483 & $W$ only, Fermi\\
 \citet{Demetriou_2002}      &4         &0.237 & $W$ and $W_s$, Fermi \\
 \citet{Demetriou_2002}      &5         &0.096 & Dispersive \\
 \citet{avrigeanu-2014}      &6         &0.182 & $W$ and $W_s$, Linear \\
 \citet{Nolte_1987}          &7         &5.403 & $W$ only. Linear\\   
 \citet{avrigeanu-1994}      &8         &2.040 & $W$ only. Linear\\
 \citet{mohr-2021}           &*        &0.237 & $W$ only. \\
\bottomrule
\end{tabular}
    \caption{Predictions of the \aomp included in TALYS and their predicted cross sections for the
    $^{86}$Sr$(\alpha, n)$ cross section at $E_{\alpha} = 6.2$ MeV. We have listed whether the 
    imaginary potential has volume ($W$) and/or surface terms ($W_S$) and its energy dependence where applicable. Note that the \aomp from \citet{mohr-2021} is calculated using a modified version of TALYS 1.8 as described in that study.}
    \label{tab:talys-model-calcs}
\end{table*}

\begin{table*}
\ra{1.3}
\begin{tabular}{@{}lllllllllllll@{}}
\toprule\toprule
TALYS ID & $\sigma_{(\alpha, n)}$ ($\mu$b) & $V$ (MeV) & $r$ (fm)& $a$ (fm)& $W$ (MeV)& $r_{i;v}$ (fm)& $a_{i;v}$ (fm)& $W_s$ (MeV)& $r_{i;s}$ (fm)& $a_{i;s}$ (fm)& $r_c$ (fm)& \\ \hline
1        & 1.002 &223.5 & 1.21   & 0.66 & 0.83  & 1.21 & 0.66 & 21.12 & 1.27 & 0.55 & 1.24 \\
2        & 0.642 &185.0 & 1.4    & 0.52 & 25.0     & 1.4     & 0.52    &     &  &  & 1.3     \\
6        & 0.182 &154.9 & 1.25   & 0.67 &      &    &      & 4.0     & 1.52    & 0.40 & 1.3     \\
7        & 5.403 &151.6 & 1.24   & 0.78 & 19.33 & 1.57    & 0.60 &     &  &  & 1.3     \\
8        & 2.040 &151.6 & 1.24   & 0.78 & 6.35  & 1.57    & 0.60 &      &  &  & 1.3     \\
\bottomrule
\end{tabular}
    \caption{TALYS' predictions for the $^{86}$Sr$(\alpha, n)$ cross section at $E_{\alpha} = 6.2$ MeV along with the corresponding 
    \aomp model parameters. The models of Ref.~\cite{Demetriou_2002} are tabulated on a grid in TALYS and are, therefore, not included
    in the table. $W_s$ is the derivative or surface imaginary potential depth using the convention $4a_{i;s}\frac{\partial{f}}{\partial{r_{i;s}}}$.}
    \label{tab:talys-model-params}
\end{table*}

\subsection{Bayesian Model}

The considerations above provide parametric forms for the \aomp model and baseline calculations to observe the impact of said parameters on { Hauser-Feshbach} model calculations. It is now necessary to relate our data from the \Sraa experiment to the \aomp and extract potential parameters and uncertainties. By doing this, we can calculate uncertainties for the \Sran cross section in light of our experimental results. Due to the complexity of this problem, we choose to use a Bayesian approach to build a statistical model that accounts for both theoretical and experimental uncertainties. 

Bayesian statistics has proven to be effective in dealing with the parametric uncertainties arising from phenomenological models commonly used in nuclear physics. Global optical model uncertainties \cite{Pruitt_2023}, transfer reactions \cite{lovell_2018, Marshall-2020, Marshall_2023}, and R-matrix analysis \cite{Odell_2022} have all successfully been analyzed using Bayesian methods and have, in most cases, provided quantified uncertainties often overlooked in frequentist analysis of similar data. However, until this work, no study has attempted to quantify \aomp uncertainties for $\alpha$-induced reactions using these techniques despite the clear need.

The heart of Bayesian statistics is using Bayes' theorem in conjunction with observations to update prior beliefs about model parameters. Bayes' theorem is given by: 
\begin{equation}
  \label{eq:bayes_theorem}
  P(\boldsymbol{\theta}|\mathbf{D}) = \frac{P(\mathbf{D}|\boldsymbol{\theta}) P(\boldsymbol{\theta})}
  {P(\mathbf{D})},
\end{equation}
where $\boldsymbol{\theta}$ are the model parameters, $\mathbf{D}$ are the data, $P(\boldsymbol{\theta})$ are the prior probability distributions which describe the model parameters before inference, $P(\mathbf{D}|\boldsymbol{\theta})$ is the likelihood function which describes the probability of parameter values given the observed data, $P(\mathbf{D})$ is the evidence (a normalization constant), and $P(\boldsymbol{\theta}|\mathbf{D})$ is the posterior \cite{sivia_2006}. The posterior distributions of interest in this work will be the \aomp parameters subject to the influence of our elastic scattering angular distributions.

\subsection{Priors For Experimental Parameters}
\label{sec:priors-exper-param}

Our priors on parameters related to experimental uncertainties follow motivations similar to those in Refs.~\cite{deSouza_2019a, deSouza_2019b, deSouza_2020, Marshall-2020, Marshall_2023}. We allow for the possibility that our data have an additional source of scatter, whether it be from experimental or theoretical sources, and that the normalization of the data is free to vary. Both of these parameters are assigned to the \textit{theoretical} predictions of the \aomp\!, a practice that we have found to improve the efficiency of sampling our model compared to scaling and spreading the \textit{experimental} data. In theory, the normalization step should be unnecessary, but as mentioned in Sec.~\ref{sec:86Sr-alpha-alpha}, the suppression voltage of the Faraday cup was incorrectly applied, meaning there is an energy dependent normalization correction. It has been claimed (i.e. in Ref.~\cite{hodgson_1994}) that leaving normalization as a free parameter in optical model calculation results in poor or unphysical fits to the data. However, both Refs.~\cite{Marshall-2020, Marshall_2023} left the normalization as a free parameter and were able to recover results consistent with independent studies and analyses. The claim is likely to stem from the difficulty of optimizing the optical model parameters, but the Bayesian method used here allows the optical model parameters to be biased towards their physical range through priors improving the behavior of the fit.

Taking an \aomp calculation at each laboratory energy, indexed $k$, and center-of-mass angle, indexed $j$, we have the normalization prior:
\begin{equation}
  \label{eq:norm-prior-g}
  g_k \sim \textnormal{Uniform}(-10, 10),
\end{equation}
where $\sim$ means ``distributed according to''. This parameter is transformed into a normalization that takes the cross section to the experimental yield:
\begin{equation}
  \label{eq:norm-parameter-func}
  \eta_k = 10^{g_k},
\end{equation}
allowing variations $\pm10$ orders of magnitude. The theoretical/predicted yield is then:
\begin{equation}
  \label{eq:theory-yield}
  \frac{dY'}{d \Omega}_{\text{Theory}, k, j} = \eta_k \frac{d \sigma}{d \Omega}_{\text{Theory}, k, j}
\end{equation}
Our additional scatter is implemented as a percentage of the theoretical yield, with a prior:
\begin{equation}
  \label{eq:scatter-prior}
  f_k \sim \textnormal{HalfNormal}(0.15^2),
\end{equation}
i.e. a half-normal distribution with a standard deviation of $15\%$ expressing out belief that it is unlikely that our data contains additional statistical scatter in excess of $15 \%$. This fractional uncertainty is added in quadrature with the measured statistical errors ($\sigma_{\textnormal{Exp}, k, j}$):
\begin{equation}
  \label{eq:total-uncertainty}
\sigma_{k, j}^{\prime 2} = \sigma_{\textnormal{Exp}, {k, j}}^2 +  \bigg(f_k\frac{d Y}{d \Omega}^{\prime}_{\textnormal{Theory}, k, j}\bigg)^2  
\end{equation}
The final experimental uncertainties considered in the model are the energy uncertainties of the $\alpha$-beam, which originate from the magnet calibration procedure (Sec.~\ref{sec:magnet-cal}). The uncertainty of each energy is taken to be the quadrature sum of the statistical and systematic errors reported in Sec.~\ref{sec:magnet-cal}. While the systematic errors are likely correlated between energies, we have taken them to be independent. The prior is a normal distribution:
\begin{equation}
  \label{eq:beam-energy-prior}
  E_{\alpha; k} \sim \mathcal{N}(E_{cal; k}, \sigma_{stat; k}^2 + \sigma_{sys; k}^2),
\end{equation}
where $E_{cal; k}$ is the $\alpha$-energy predicted by our magnet calibration at energy $k$.

\subsection{\aomp Priors}
\label{sec:aomp-priors}

For the parameters of the Woods-Saxon volume potential, we center our priors around the values of McFadden and Satchler \cite{McFadden-1966} \footnotemark[1]. Standard deviations taken to be $20 \%$ of the central value in order to comfortably cover the physical range of these parameters ($r=[1.1, 1.7]$ and $a = [0.42, 0.62]$). We have:
\begin{equation}
  \label{eq:aomp-priors}
  \mathcal{U} \sim \mathcal{N}(\mu_{\textnormal{M\&F}}, [0.2 \mu_{\textnormal{M\&F}}]^2),
\end{equation}
where $\mu_{\textnormal{M\&F}}$ refers to the parameters of the McFadden and Satchler optical model.

\footnotetext[1]{It should be noted, that while the literature frequently refers to the McFadden and Satchler potential parameters as a global model, in the words of the author's of that study they are only a ``...convenient starting set in a search for potentials'' \cite{McFadden-1966}, which is exactly the purpose they serve in our study.}

For the energy dependence, we have two sets of priors for the linear and Fermi parametrization, respectively. All of the priors for the energy dependence were constructed to overlap with the global trends or local values reported in Refs.~\cite{avrigeanu-1994, Somorjai-1998} and to roughly align with the McFadden and Satchler values at $E_{\alpha} = 24.7$ MeV.

The real potential energy dependence is taken to be linear and decreasing with energy (Eq.~(\ref{eq:real-depth-linear})). The parameter $V_0$ is a normal distribution centered around the McFadden and Satchler value of $150$ MeV with a standard deviation of $20 \%$. The parameter $V_1$ is distributed according to a half-normal distribution with mean of 0 and a variance of 1, ensuring that the sign of the dependence does not change while covering the ranges typical of other studies:
\begin{equation}
  \label{eq:real-energy-dep-prior}
  V_1 \sim \textnormal{HalfNormal}(1).
\end{equation}
The linear dependence for the imaginary volume depth is the same except increasing (Eq.~(\ref{eq:imaginary-depth-linear})) with energy. It was found that the $20 \%$ standard deviation centered around the McFadden and Satchler value of $25$ MeV was too constrained for this case. To remedy this, $W_0$ was allowed a $40 \%$ spread around the $25$ MeV depth:
\begin{equation}
  \label{eq:img-linear-energy-w0}
  W_0 \sim \mathcal{N}(25.0, [0.4(25.0)]^2).
\end{equation}
The slope, $W_1$, follows the same prior as the real slope:
\begin{equation}
  \label{eq:img-linear-energy-dep-prior}
  W_1 \sim \textnormal{HalfNormal}(1),
\end{equation}
The Fermi form has three free parameters, taken to be:
\begin{equation}
  \label{eq:img-fermi-zero}
  W_0 \sim \mathcal{N}(25.0, [0.2(25.0)]^2),
\end{equation}
\begin{equation}
  \label{eq:img-fermi-one}
  W_1 \sim T(0, \infty)\mathcal{N}(10, 0.5^2),
\end{equation}
and
\begin{equation}
  \label{eq:img-fermi-two}
  W_2 \sim T(0, \infty)\mathcal{N}(2.0, 0.5^2),
\end{equation}
with $T(0, \infty)\mathcal{N}$ being the truncated normal distribution restricted to positive values. The range for these parameters was a rough estimate of their acceptable physical values.

\subsection{Constraint on the Discrete Ambiguity}
\label{sec:constr-discr-ambig}

Elastic scattering cross sections calculated from the optical model are nearly degenerate when the phase shift differs by a multiple of $\pi$. The real potential depth can produce these shifts related by $n \pi$, resulting in a multi-modal posterior distribution, commonly referred to as the ``discrete ambiguity''. This behavior is severe for a strongly absorbed projectile like $^{4}$He \cite{drisko_1963}, and our initial attempts to sample the posterior resulted in families spanning $V \approx 60\text{-}300$ MeV. We adjusted our approach to allow multiple modes in our posterior distribution, but to limit those significantly different from the McFadden and Satchler parameters. We follow the method outlined in Ref.~\cite{Marshall_2023} by putting a \textit{constraint} on the posterior distribution via the real volume depth and radius \footnotemark[1]. We define a quantity, $c$, such that:
\begin{equation}
  \label{eq:vr-constant}
  c = Vr^{n},
\end{equation}
where $n$ is a free parameter. In order to select a value for $n$, we used samples from the unconstrained posterior (a detailed description of the sampling can be found later in the article in Sec.~\ref{sec:numer-cons}), constructed histograms for $c$, and then varied $n$ by hand until the separation between modes appeared to be maximized. We found $n = 1.6$ to provide good separation. Restricting our sampling to $380 < c < 270$ MeV$\cdot$fm$^n$ gave us the modes closest to the global value ($c = 259$).
\footnotetext[1]{The relationship between the real radius and depth is often referred to as the  ``continuous ambiguity'', but is nothing more than run-of-the-mill parameter correlation.}

\subsection{Bayesian Model Summary}
\label{sec:bayes-model-summ}
We have detailed the Bayesian model constructed to both fit the \Sraa data and then predict the \Sran cross section. For clarity we will write the model in its entirety and reiterate the main points:
\begin{enumerate}
\item We treat the data from each energy as having an unknown normalization and additional statistical uncertainty.
\item We constrain the posterior modes that arise due the discrete ambiguity via Eq.~(\ref{eq:vr-constant}).
\item Our potential priors are centered around the values of McFadden and Satchler \cite{McFadden-1966}.
\item Two separate models are constructed and fit that assume either a linear or Fermi energy dependence for the imaginary depth.
\end{enumerate}

\begin{align}
  \label{eq:bayesian-model}
 & \textnormal{\textbf{Priors:}} \nonumber \\
 & E_{\alpha; k} \sim \mathcal{N}(E_{cal; k}, \sigma_{stat; k}^2 + \sigma_{sys; k}^2) \nonumber \\
 & V_0 \sim \mathcal{N}(185, [0.2(185)]^2) \nonumber \\
 & V_1 \sim \textnormal{HalfNormal}(1) \nonumber \\
 & r \sim \mathcal{N}(1.4, [0.2(1.4)]^2) \nonumber \\
 & a \sim \mathcal{N}(0.52, [0.2(0.52)]^2) \nonumber \\ \cline{1-2}
 & \textnormal{Linear:} \nonumber \\
 & W_0 \sim \mathcal{N}(25, [0.4(25)]^2) \nonumber \\
 & W_1 \sim \textnormal{HalfNormal}(1) \nonumber \\ \cline{1-2}
 & \textnormal{Fermi:} \nonumber \\
 & W_0 \sim \mathcal{N}(25, [0.2(25)]^2) \nonumber \\
 & W_1 \sim T(0, \infty)\mathcal{N}(10, 0.5^2) \nonumber \\
 & W_2 \sim T(0, \infty)\mathcal{N}(2.0, 0.5^2) \nonumber \\ \cline{1-2}
 & r_i \sim \mathcal{N}(1.4, [0.2(1.4)]^2) \nonumber \\
 & a_i \sim \mathcal{N}(0.52, [0.2(0.52)]^2) \nonumber \\
 & g_k \sim  \textnormal{Uniform}(-10.0, 10.0) \nonumber \\
 & f_k \sim \textnormal{HalfNormal}(0.15) \nonumber \\
 & \textnormal{\textbf{Functions:}} \\
 & \eta_k = 10^{g_k}  \nonumber \\
 & V_k = V_0 - V_1 E_{\alpha; k} \nonumber \\ \cline{1-2}
 & \textnormal{Linear:} \nonumber \\
 & W_k = W_0 + W_1 E_{\alpha; k} \nonumber \\ \cline{1-2}
 & \textnormal{Fermi:} \nonumber \\
 & W_k = W_0 \bigg(1 + \exp{\frac{W_1 - E_{\alpha; k}}{W_2}} \bigg) \nonumber \\ \cline{1-2}
 & \frac{d Y}{d \Omega}^{\prime}_{\textnormal{Optical}, k, j} = \eta_k \times \frac{d \sigma}{d \Omega}_{\textnormal{Optical}, k, j} \nonumber \\
 & \sigma_{k, j}^{\prime 2} = \sigma_{\textnormal{Exp}, {k, j}}^2 +  \bigg(f_k\frac{d Y}{d \Omega}^{\prime}_{\textnormal{Optical}, k, j}\bigg)^2 \nonumber \\
 & \textnormal{\textbf{Likelihoods:}} \nonumber \\
 & \frac{d Y}{d \Omega}_{\textnormal{Elastic}, k, j} \sim \mathcal{N}\bigg(\frac{d Y}{d \Omega}^{\prime}_{\textnormal{Optical}, j, k}, \sigma_{\textnormal{Elastic}, i}^{\prime 2} \bigg) \nonumber \\
 & \textnormal{\textbf{Constraints:}} \nonumber \\
 & n = 1.6 \nonumber \\
 & c = V_0 r^n \nonumber \\
 & 380 < c < 270, \nonumber
\end{align}
where the index $j$ refers to the center-of-mass angle and $k$ denotes each of the laboratory energies.



\subsection{Numerical Considerations}
\label{sec:numer-cons}

Sampling of the posterior distribution was carried out using dynamic nested sampling as implemented in the \texttt{python} package \texttt{dynesty}
\cite{skilling_2004,skilling_2006,speagle_2020_dynesty,sergey_koposov_2024,Higson_2019}. Dynamic nested sampling was first introduced in Ref.~\cite{Higson_2019} as a method of improving the posterior estimations produced by traditional nested sampling. To expand on this, traditional nested sampling works by successive updating of samples drawn from the priors, called \textit{live points}, such that, with each update, the likelihood increases. These samples are drawn from the posterior, but they do not guarantee accurate parameter estimation since they are not focused around the maximum of the posterior. Dynamic nested sampling allocates additional live points in specific regions in order to decrease the variance of the parameter estimates. Users of dynamic nested sampling can set whether they want to focus on improving parameter or evidence estimates via a parameter $G = [0, 1]$, with $G=0$ providing focus on the evidence and $G=1$ on parameter estimates. Estimates of the evidence are primarily useful for the Bayesian model comparison (for an example, see the determination of $\ell$ values from transfer reactions in Ref.~\cite{Marshall-2020}). Dynamic nested sampling was preferred over Markov Chain Monte Carlo in the current work since it can more readily deal with the multi-modal posteriors produced by the discrete ambiguity.

For this paper all dynamic nested sampling runs were initialized with 500 live
points bounded with multiple ellipsoids with updates performed via slice sampling. Our focus was on parameter estimation and, as a result, $G = 1$. The algorithm required around $4.5 \times 10^{7}$ likelihood evaluations per run, corresponding to $1.3 \times 10^{8}$ \aomp calculations, and produced at least $1 \times 10^{5}$ posterior samples. A basic optical model code\footnotemark[1] was developed to render these number of evaluations tractable, and its validity was checked against the reaction code ECIS97 \cite{raynal_1971} in the energy range under consideration. It was found to agree to $3 \%$.

\footnotetext[1]{\url{https://github.com/camarsha/dwuckman}}

\section{Results and Discussion}
\label{sec:results}

\subsection{Elastic Scattering Posterior Distributions}
\label{sec:fits-post-distr}

\begin{figure}
    \centering
    \includegraphics[width=\linewidth]{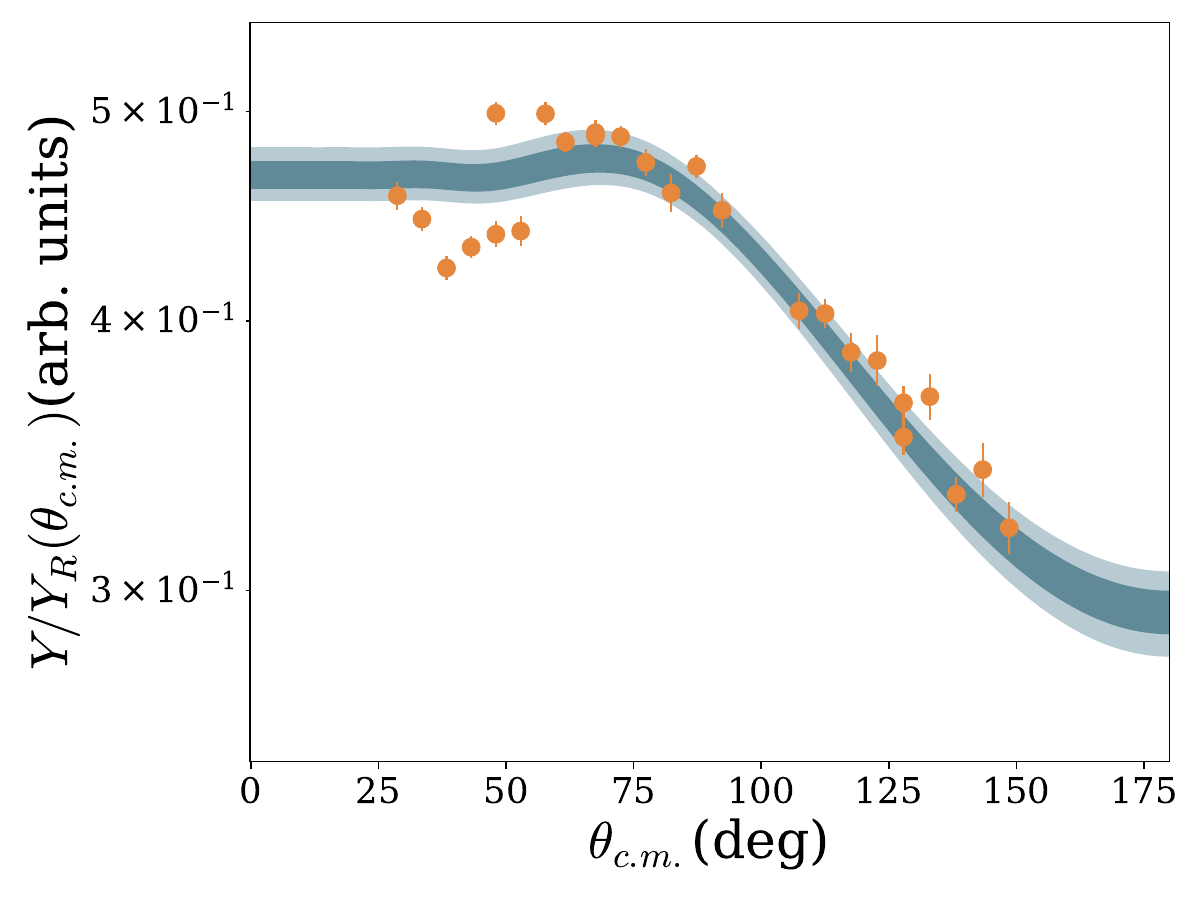}
    \caption{Angular distribution for the $12$-MeV data. The $68\%$ and $95\%$ credibility intervals from our linear optical model are given in blue and light blue, respectively.}
    \label{fig:12mev-fit}
\end{figure}

\begin{figure}
    \centering
    \includegraphics[width=\linewidth]{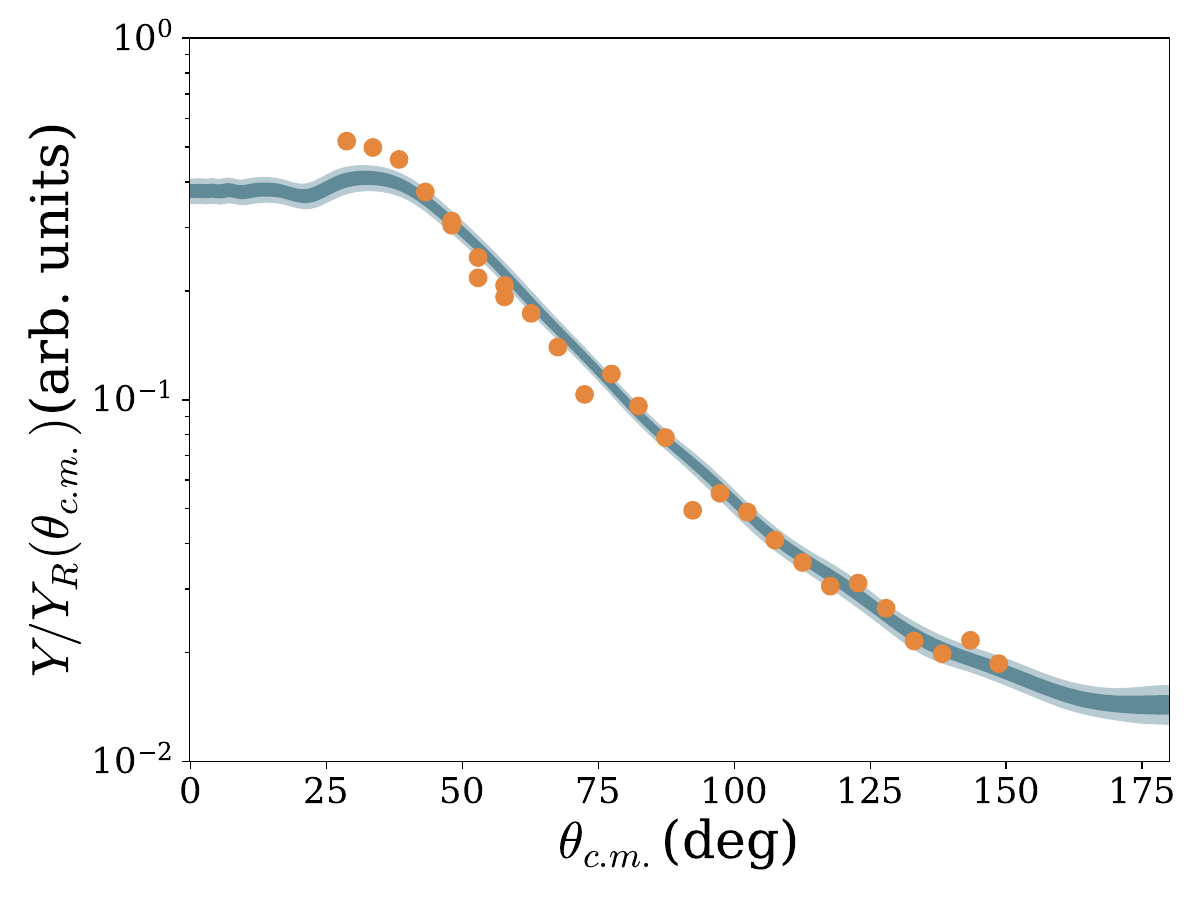}
    \caption{Angular distribution for the $18$-MeV data. The $68\%$ and $95\%$ credibility intervals from our linear optical model are given in blue and light blue, respectively.}
    \label{fig:18mev-fit}
\end{figure}

\begin{figure}
    \centering
    \includegraphics[width=\linewidth]{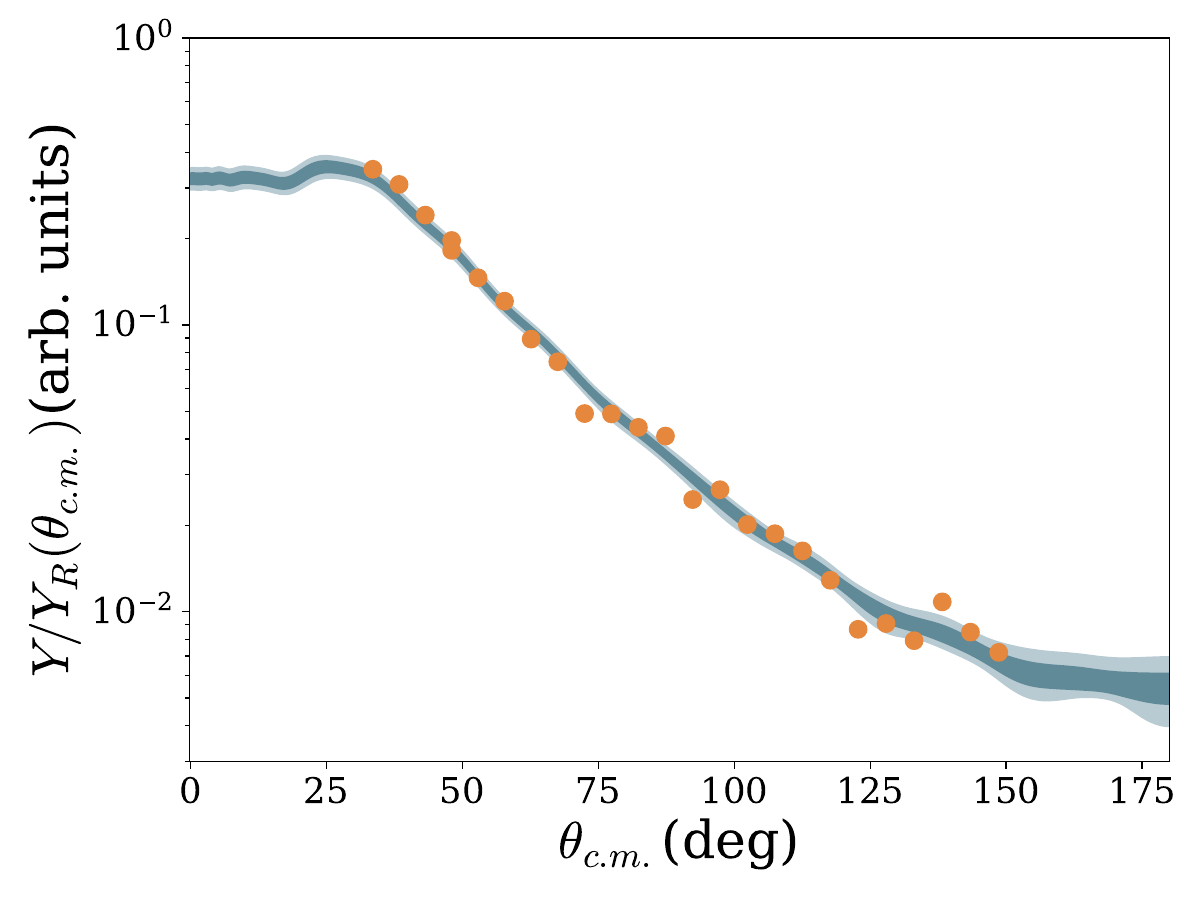}
    \caption{Angular distribution for the $21$-MeV data. The $68\%$ and $95\%$ credibility intervals from our linear optical model are given in blue and light blue, respectively.}
    \label{fig:21mev-fit}
\end{figure}

Our fits of the experimental \Sraa angular distributions are presented in Figs.~\ref{fig:12mev-fit}, \!\ref{fig:18mev-fit}, and \ref{fig:21mev-fit} along with the $68 \%$ and $95 \%$ credibility intervals of the optical model calculations. Both the Fermi and linear models produced seemingly identical distributions and, as a result, we have only shown the fits from the linear model. Note that the error bars show the statistical errors from the peak areas only and do not include the additional uncertainty estimated from the fitting procedure. Two points, $97^{\circ}$ and $102^{\circ}$, were excluded from the 12 MeV fit due to abnormally low yields.

Our linear and Fermi models have 17 and 18 parameters, respectively. Samples produced from nested sampling allow us to examine correlations between all of these parameters and compare their behaviour between the two models. We divide the parameters into the following subgroups: experimental, real potential, and imaginary potential. Each parameter's posterior distribution will be compared to its prior one; the more similar these distributions are, the less our data has influenced their final value and uncertainty.

\begin{figure*}
  \centering
  \includegraphics[width=\linewidth]{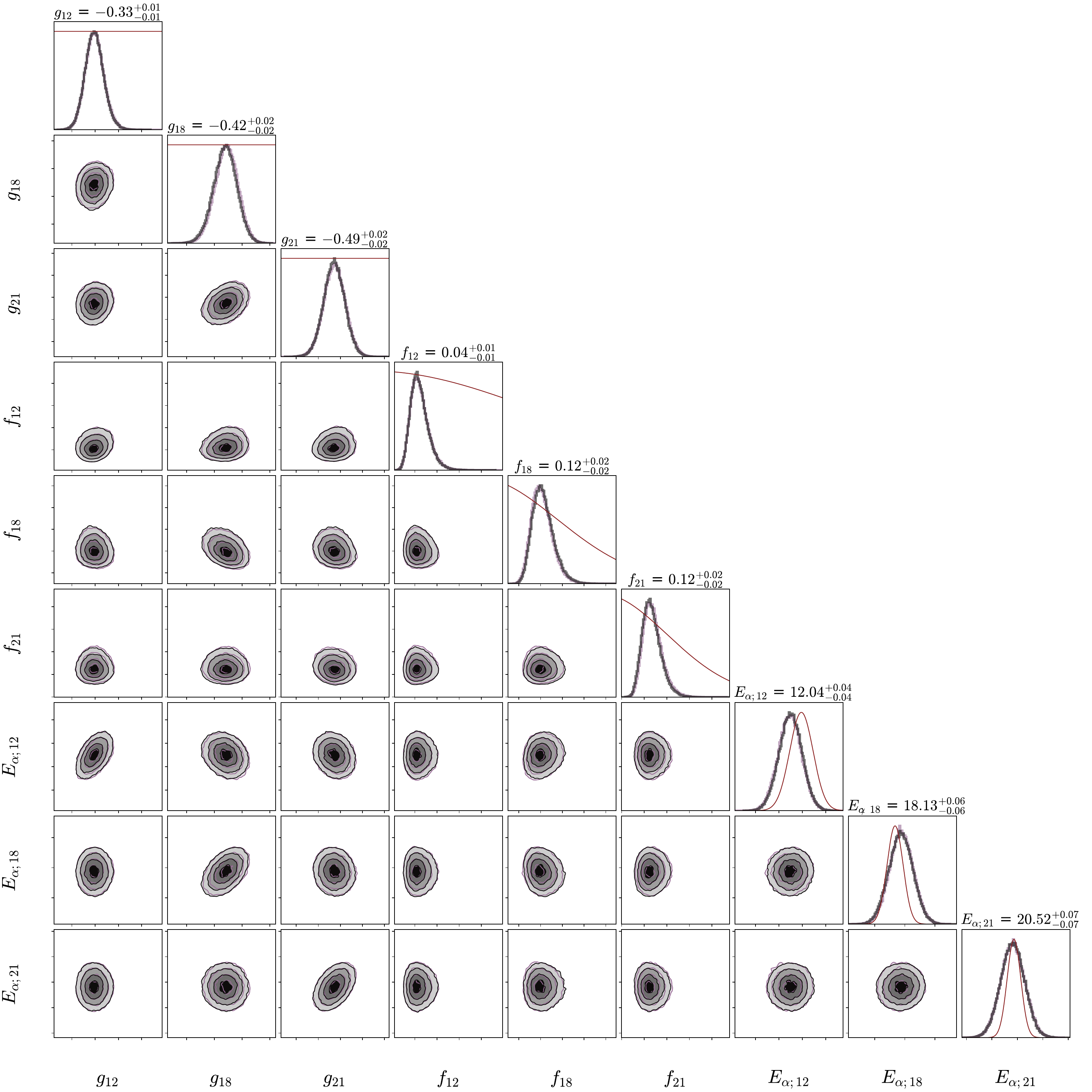}
  \caption{Corner plot for the experimental parameters: overall normalization ($g$), additional scatter ($f$), and beam energy ($E_{\alpha}$). The linear and Fermi parameters are plotted in light purple and black, respectively, while the prior distributions are shown as a solid red line in the 1D histograms. The linear and Fermi contours overlap nearly exactly. Reported numbers are the $16$, $50$, $84$ percentiles of the linear models' posteriors. Subscripts indicate the energy of the data set they correspond to, i.e. $g_{12}$ is the normalization parameter for the $12$-MeV data set. All symbols are defined in the text, see Sec.~\ref{sec:bayes-model-summ}}
  \label{fig:corner-exp}
\end{figure*}

The experimental parameters cover the overall normalization (e.g. $g_{12}$), additional scatter (e.g. $f_{12}$), and beam energy (e.g. $E_{\alpha;12}$). Fig.~\ref{fig:corner-exp} shows the pair correlation plots for both the linear and Fermi models, with the overlap between  the two being almost exact. The overall normalization at each energy is most highly correlated with the beam energy, but amounts to less than $5 \%$ uncertainty on the cross section for all energies. The normalization also has the effect of pulling the assumed beam energies away from their prior values.


\begin{figure}
  \centering
  \includegraphics[width=\linewidth]{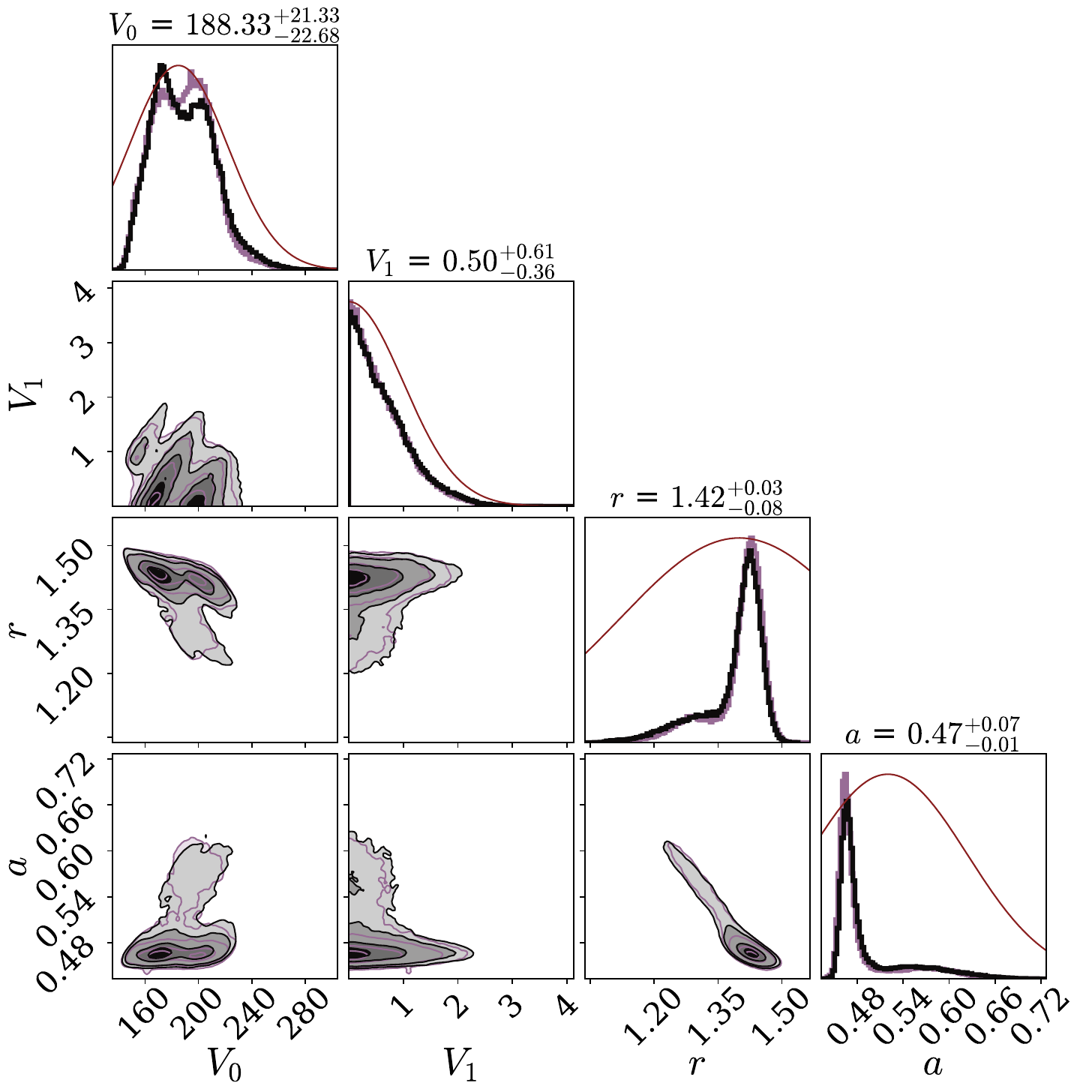}
  \caption{Corner plot for the real potential parameters. The linear and Fermi parameters are plotted in light purple and black, respectively, while the prior distributions are shown as solid red lines in the 1D histograms. Reported numbers are the $16$, $50$, $84$ percentiles of the linear models posteriors. All symbols are defined in the text, see Sec.~\ref{sec:bayes-model-summ}}
  \label{fig:corner-real}
\end{figure}

The correlations between the real parameters can be found in Fig.~\ref{fig:corner-real} for both models. Overall the differences are marginal, with the geometric parameters ($r$ and $a$) being particularly consistent. The energy dependence of the real potential ($V_1$) has deviated from its prior value, with more weight towards lower values than would be expected from a half-normal distribution with $\sigma = 1$. However, it is not clear that our data adds any particular constraint, rather it appears to be an indirect consequence of the correlation between $V$ and $r$ and the limits placed on the discrete ambiguity.

\begin{figure}
  \centering
  \includegraphics[width=\linewidth]{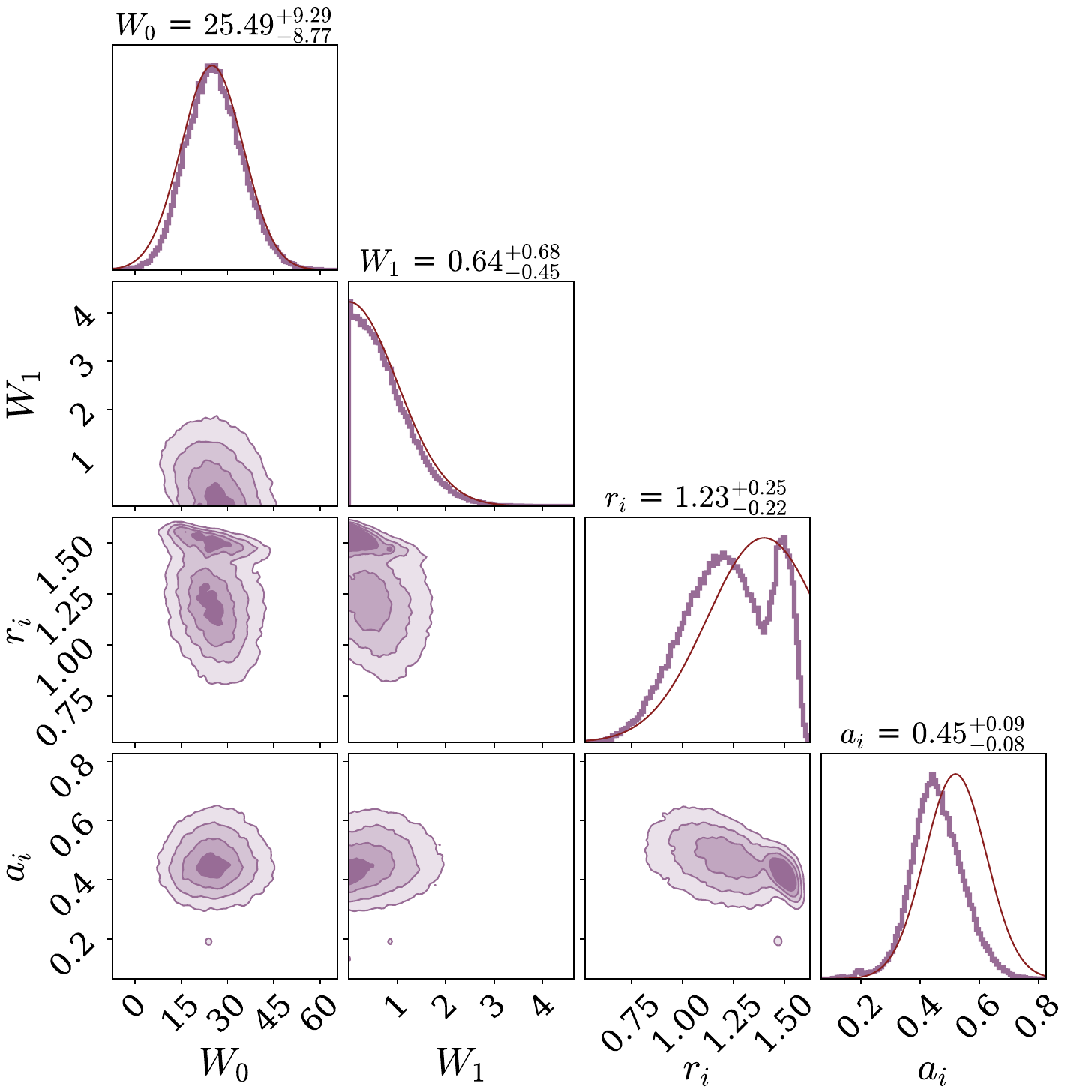}
  \caption{Corner plot for the linear imaginary potential parameters. The prior distributions are shown as solid red lines in the 1D histograms. Reported numbers are the $16$, $50$, $84$ percentiles of the posteriors. All symbols are defined in the text, see Sec.~\ref{sec:bayes-model-summ}}
  \label{fig:corner-img-1}
\end{figure}

\begin{figure}
  \centering
  \includegraphics[width=\linewidth]{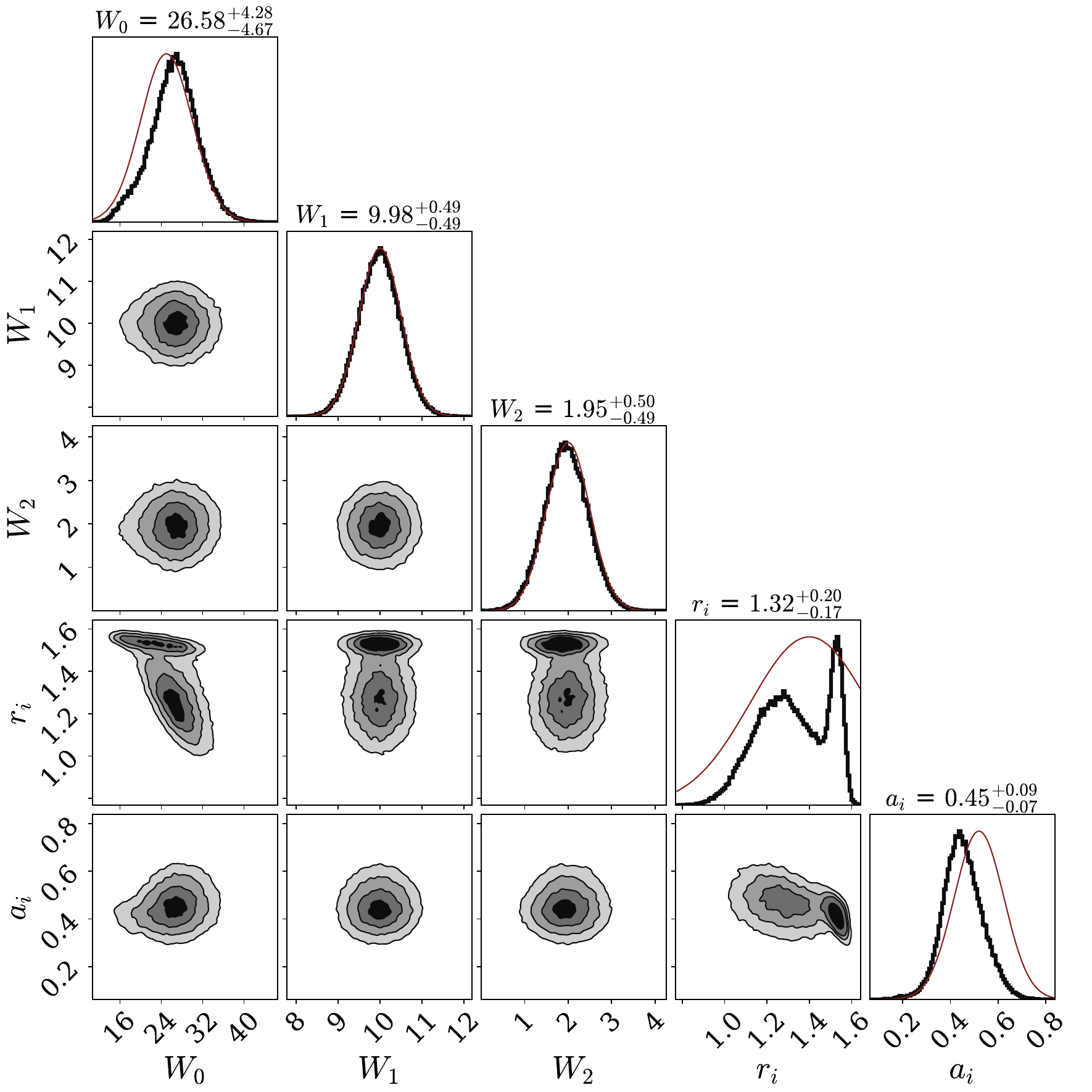}
  \caption{Corner plot for the Fermi imaginary potential parameters. The prior distributions are shown as solid red lines in the 1D histograms. Reported numbers are the $16$, $50$, $84$ percentiles of the posteriors. All symbols are defined in the text, see Sec.~\ref{sec:bayes-model-summ}}
  \label{fig:corner-img-2}
\end{figure}

The imaginary potential parameters of Figs.~\ref{fig:corner-img-1},~\ref{fig:corner-img-2} indicate similar behavior to the real parameters, but with even less of a deviation from their prior values. In particular, the parameters that control the energy dependence for the linear and Fermi models are unchanged from their prior values. Our data does not constrain the energy dependence of the imaginary potential at all, and the posterior value is only based on our initial assumptions. The geometric parameters are significantly more constrained than the potential strengths, but still less sharply peaked than their real counterparts.
 
\subsection{Cross Section Prediction at $E_{\alpha} = 6.2$ MeV}
\label{sec:cross-sect-pred}

\begin{figure*}
    \centering
    \includegraphics[width=\linewidth]{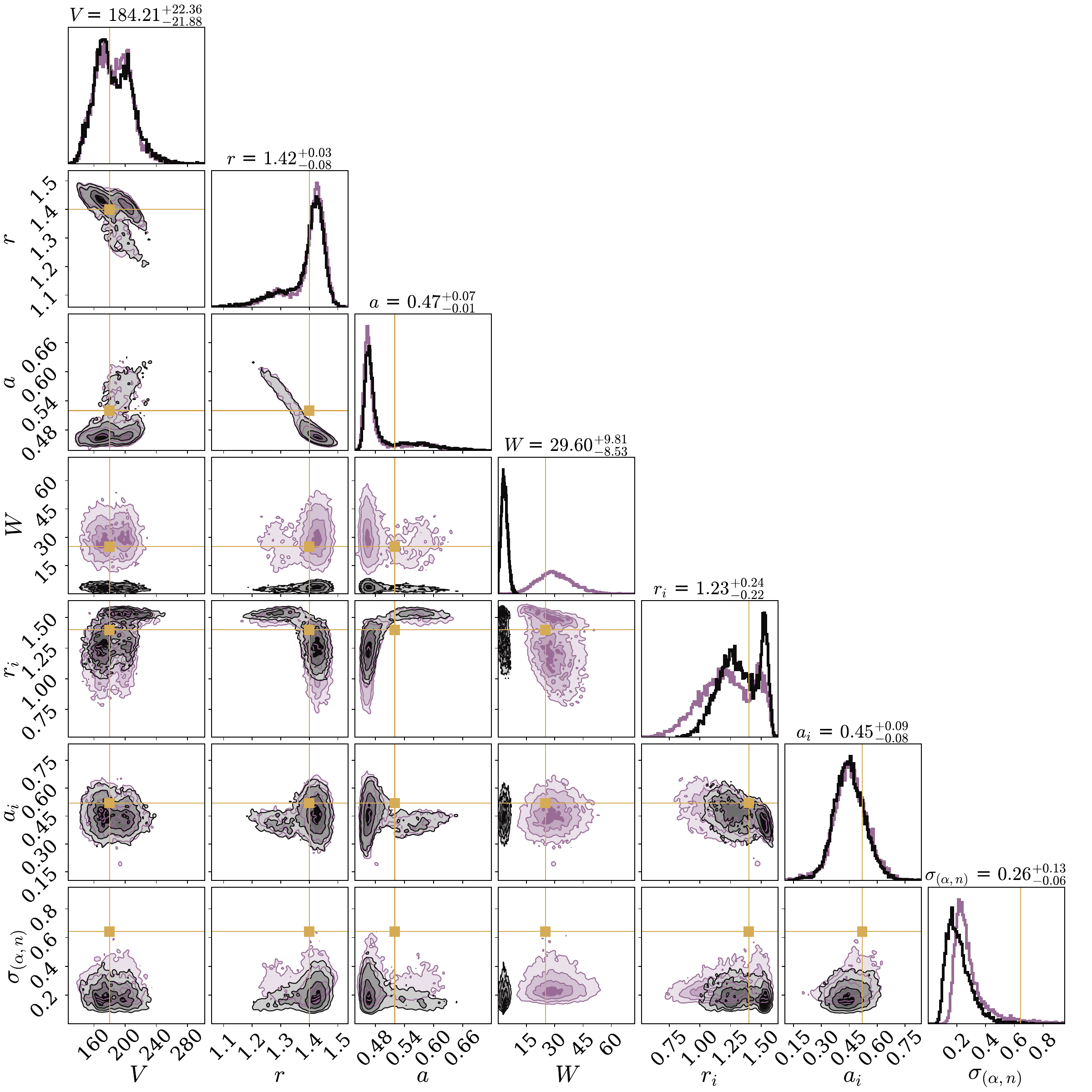}
    \caption{Pair correlation plot of the \Sran cross sections at $E_{\alpha} = 6.2$ MeV and the potential parameters. The predictions of the linear energy dependence model are in light purple, while those of the Fermi model are in black. See text for additional details. The parameters of McFadden and Satchler are shown with the yellow lines.}
    \label{fig:hf-comp-corner}
\end{figure*}

We use the posterior samples shown in the previous section to construct $6$-parameter \aomp samples for TALYS calculations at $E_{\alpha}^{\text{lab.}} = 6.2$ MeV using both the linear and Fermi models. Due to computational considerations, we only use $1 \times 10^4$ samples for each model. Fig.~\ref{fig:hf-comp-corner} shows the pair correlation plot between the potential parameters and the \Sran cross section. Considering the radically different $W$ values for the linear and Fermi models, it is remarkable that the \Sran cross section distributions overlap significantly. No strong correlations are visually observed between the cross section and the potential parameters, with perhaps the exception of the imaginary diffuseness parameter. To try and quantify this correlation beyond visual inspection, we use Spearman's rank order coefficient, $r_{spearman}$, which quantifies correlations between -1 and 1 \cite{stat_methods}. The only variables that have an appreciable correlation, $|r_{spearman}| > 0.1$, with the cross section for the linear case are: $a$ with $0.20$, $r_i$ with $0.23$, and $a_i$ with $0.47$. For the Fermi case we find: $r$ with $0.24$, $a$ with $0.15$, and $a_i$ with $0.16$. It is our interpretation that these correlations with the cross section reflect the strong correlations between the geometric parameters of the real and imaginary potentials.

Going further, the median cross sections of both models are significantly displaced from the predictions of the McFadden and Satchler parameters that we based our priors on (the yellow lines in Fig.~\ref{fig:hf-comp-corner}). It is noteworthy that there is seeming agreement between our optical model parameters and those of McFadden and Satchler when just comparing the parameters individually. Our posterior samples, however, indicate that in the high dimensional parameter space of the \aomp\!, the McFadden and Satchler parameters actually lie in the tails of the posterior distribution. These observations imply that the cross section predictions of the \aomp are not sensitive to any one parameter and are instead the result of the highly correlated parameter space.

\begin{figure}
    \centering
    \includegraphics[width=\linewidth]{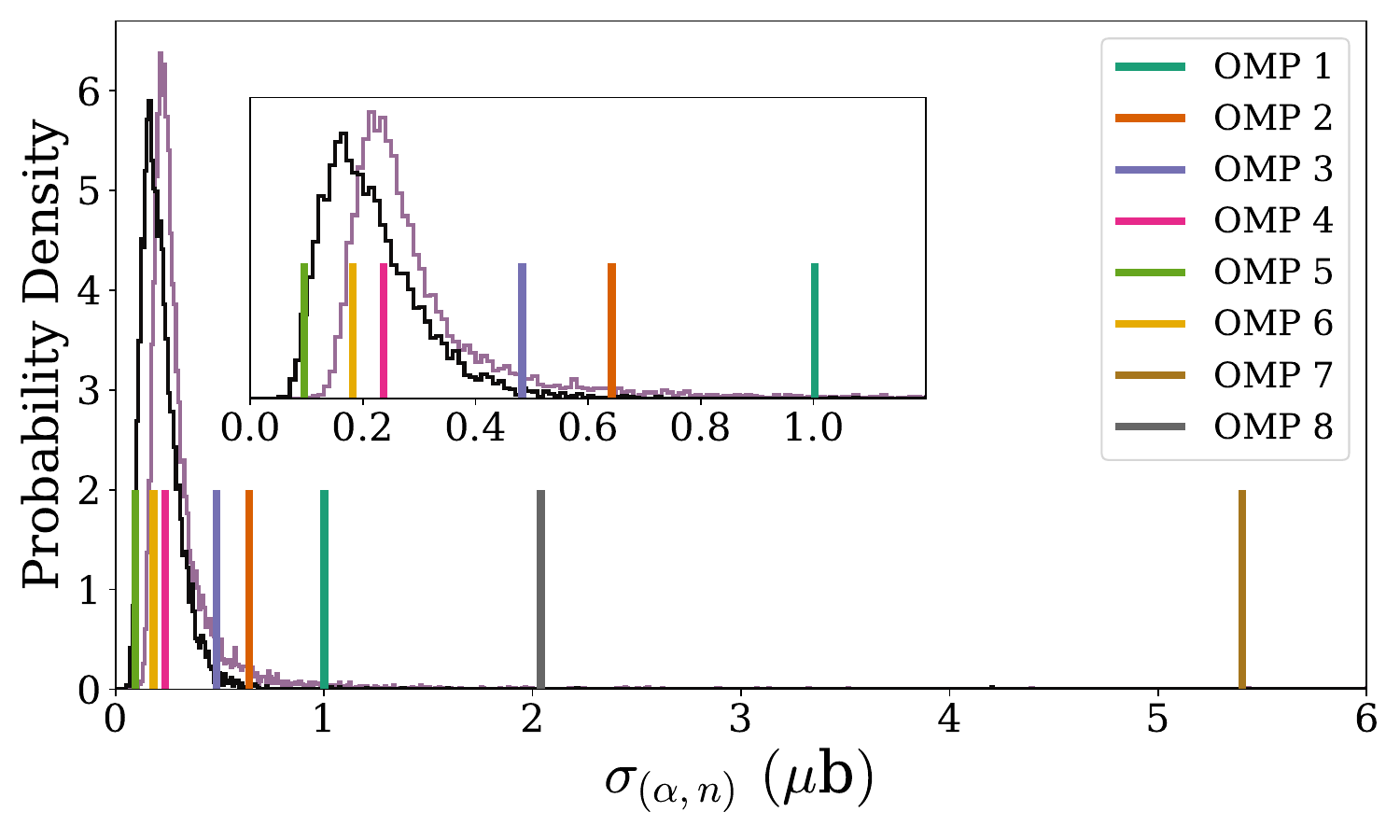}
    \caption{Comparison of our predicted \Sran cross sections at $E_{\alpha} = 6.2$ MeV and the predictions of the \aomp potentials included within TALYS. The predictions of the linear energy dependence model are in light purple, while those of the Fermi model are in black. See text for additional details.}
    \label{fig:hf-comp-all-omp}
\end{figure}

The \Sran cross sections can also be compared to all of \aomp models provided in TALYS (see Tables \ref{tab:talys-model-calcs} and \ref{tab:talys-model-params}.) Fig.~\ref{fig:hf-comp-all-omp} shows that, in general, we overlap with the majority of the provided models, but disfavor the predictions of \citet{avrigeanu-1994},  \citet{Nolte_1987}, and the TALYS potential constructed from single nucleon potentials.
We give our suggested values and $68 \%$ credibility intervals for the \Sran cross section at $E_{\alpha} = 6.2$ MeV in Table \ref{tab:cs-pred}. The ratio of the upper ($84 \%$) and lower ($16\%$) percentiles with the median are also provided to give a rough idea of factor uncertainty. It is from these ratios that we draw the conclusion that our elastic scattering data are capable of constraining the cross section to $50 \%$ at astrophysical energies.

\begin{table}
  \ra{1.3}
  \centering
  \begin{tabular}{llll}
    \toprule\toprule
    Model & $\sigma_{(\alpha, n)}$ ($\mu$b) & $84\% / 50 \%$ & $50 \% / 16 \%$ \\ \hline
    Linear & $0.26^{+0.13}_{-0.06}$ & $1.5$  & $1.3$ \\
    Fermi & $0.20^{+0.10}_{-0.06}$ & $1.5$ & $1.43$ \\
    Linear + \citet{oprea-2017} & $0.51^{+0.8}_{-0.14}$ & $2.57$ & $1.38$ \\
    \bottomrule
  \end{tabular}
  \caption{Cross sections with $68 \%$ credibility intervals at $E_{\alpha} = 6.2$ MeV. The ratios of the upper ($84 \%$) and lower ($16\%$) percentiles with the median are given to show the approximate factor uncertainty of the cross section predictions.}
  \label{tab:cs-pred}
\end{table}

\begin{figure}
  \centering
  \includegraphics[width=\linewidth]{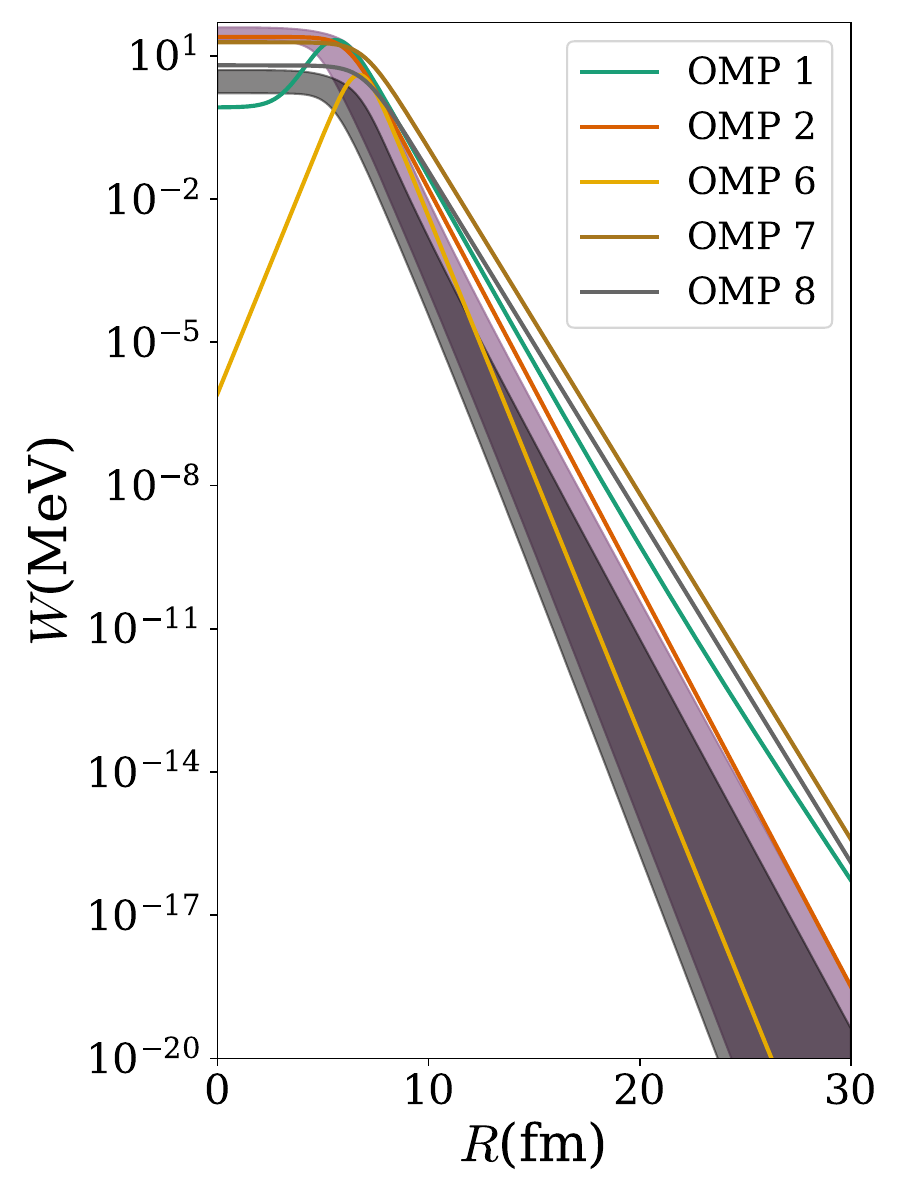}
  \caption{The total imaginary potential strength as a function of radius. The purple and black bands correspond to the $68 \%$ credibility intervals for the potential strength for the linear and Fermi models, reflectivity. The solid lines show the strengths for the TALYS parameters found in Table~\ref{tab:talys-model-params}. There is a close correspondence between the overall imaginary strength and the predicted \Sran cross sections for radii between $15\text{-}20$ fm.}
  \label{fig:w-vs-r}
\end{figure}

As a final investigation, we look for an origin for the spread in cross section values.
It has been known since the late 50s that the $\alpha$-elastic scattering cross section is sensitive only to the surface of the Woods-Saxon potential \cite{Igo_1958,Igo_1959}. The continuous ambiguity arises from elastic scattering being primarily influenced by scattering at the nuclear surface where the potential strength is a function, not just of the depth, but also the geometric parameters. Our two models for the potential depth's energy dependence support this view, since each model has significantly different imaginary depths yet largely consistent cross sections. Despite this, we did not find evidence for the relationship found in Refs.~\cite{Igo_1959,drisko_1963} where ``equivalent'' potentials meet the condition $C = W \exp(rA^{1/3}/a)$, where $C$ is a constant. Our posterior samples demonstrated scatter of many orders of magnitude for $C$ and no obvious trend, which could be the result of the diffuseness being allowed to vary \cite{Jackson_1968}. Another view comes from Mohr et. al \cite{Mohr_2020}, where a sensitivity of what these authors call the ``tail'' of the imaginary potential was found (see also Ref.~\cite{Frahn_1963}). They truncated the imaginary potential at 12 fm and observing large changes in $\sigma_{reac}$ for $^{197}$Au$+ \alpha$. Calculating $Wf(r;r_i,a_i)$ for $r=12$ fm, we observed a similar relationship as that was observed in the cross section data in Fig.~\ref{fig:hf-comp-all-omp}. Pursuing this explanation further, we constructed the creditability intervals for $Wf(r;r_i,a_i)$ from $r=0\text{-}30$ fm and plotted them against the other Models in TALYS for which we have analytical forms (see Table~\ref{tab:talys-model-params}). The results can be found in Fig.~\ref{fig:w-vs-r}. Comparing Figs.~\ref{fig:hf-comp-all-omp} and \ref{fig:w-vs-r}, we can see a clear correspondence with the behavior of the imaginary potential strength at $10-20$ fm and the \Sran cross sections.




\subsection{Inclusion of Cross Section Data}
\label{sec:potential-impacts-cross}

So far, we have shown that our \Sraa data did little to constrain the energy dependence of the \aomp parameters despite the measurements at multiple energies. Contrary to expectations, this does not hinder our ability to make relatively precise predictions of the reaction cross section at lower energies. However, we do observe that our cross section has been pulled away from the predictions of the McFadden and Satchler parameters, and that these parameters have been shown to agree with the cross section measurements of Ref.~\cite{oprea-2017}. If the McFadden and Satchler predictions are accurate, then it could be the case that the optical model uncertainties are not dominant, but rather that there is a systematic deficiency with our models that produce precise but inaccurate extrapolations at low energies.

To investigate, we repeated the analysis of the linear energy dependence model, but included the data of Ref.~\cite{oprea-2017}. At the three energies reported in that study, 10.18, 10.79, and 11.40 MeV TALYS predicts a relatively constant neutron branching ratio of $\approx 93 \%$, with $\sigma_{(\alpha, p)} + \sigma_{(\alpha, n)} \approx \sigma_{reac}$. { The branching ratio at these energies was also found to be significantly less sensitive to the level density than that at $6.2$ MeV (Sec.~\ref{sec:stat-model}).}  The total reaction cross section can be calculated using the phase shifts, $\delta_j$, of the optical potential using:

\begin{equation}
  \label{eq:total-reac-cross}
  \sigma_{reac} = \frac{4 \pi}{k^2} \sum_{\ell} (2 \ell + 1) [\mathrm{Im}(C_{\ell}) - |C_{\ell}|^2], 
\end{equation}
with $C_{\ell} = -i/2[\exp(2i\delta_{\ell}) - 1]$ \cite{melkanoff_1965}.
To compare to the \Sraa data, we scale $\sigma_{reac}$ by 0.93 and assume a $10 \%$ uncertainty on the resulting prediction. In this way, we can efficiently calculate the cross sections at these energies making inference with these data feasible. Our Bayesian model was updated to include an additional likelihood term:

\begin{equation}
  \label{eq:cs_likelihood}
  \sigma_{exp; m} \sim \mathcal{N}(0.93\sigma_{reac; m} , [0.093\sigma_{reac; m}]^2),
\end{equation}
for each energy, $m$. Inference was carried out in an identical fashion to all other calculations presented here using the linear model.

\begin{figure}
    \centering
    \includegraphics[width=\linewidth]{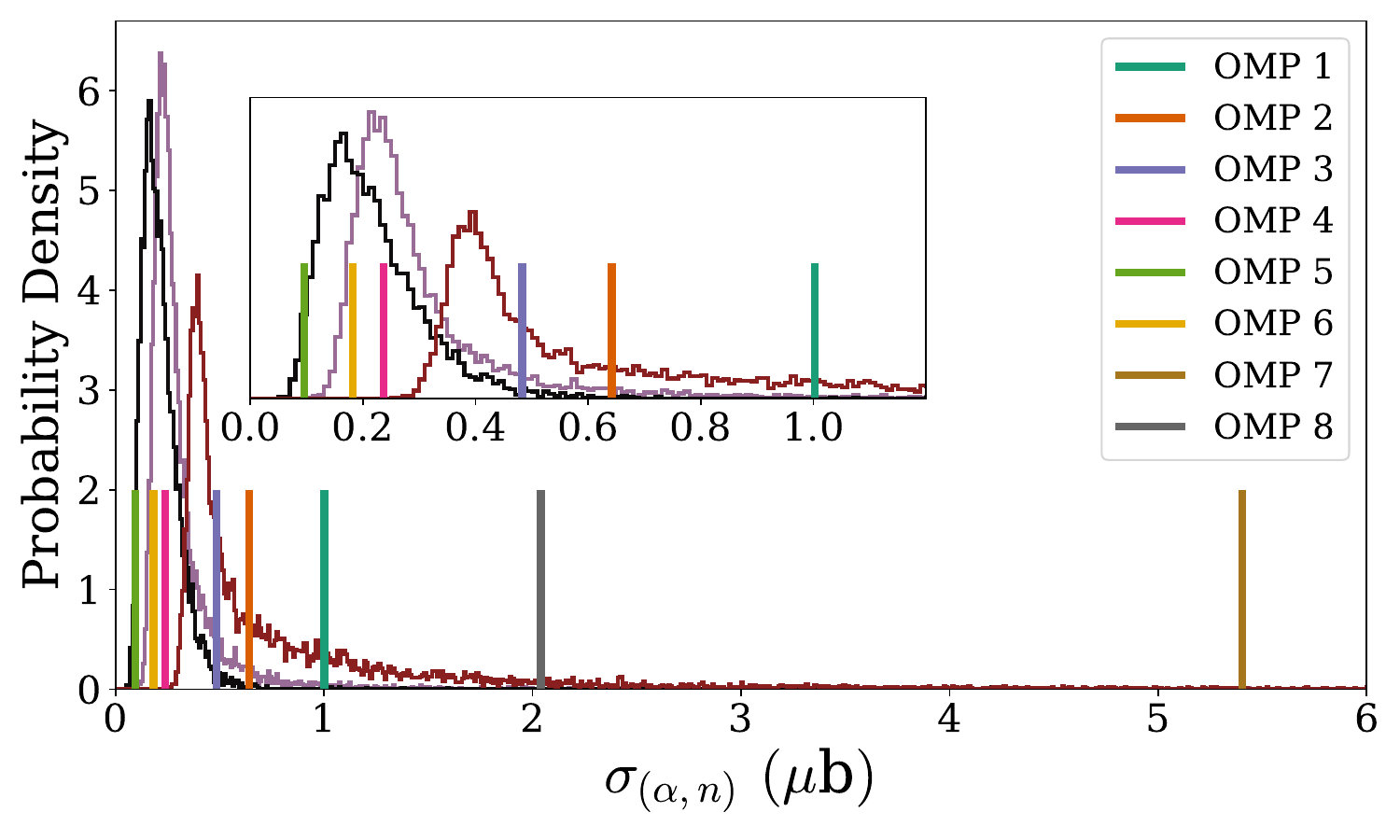}
    \caption{Same as Fig.~\ref{fig:hf-comp-all-omp}, but with the additional predictions of \Sran when including in the cross section data of Ref.~\cite{oprea-2017} in the linear model. The posterior samples of the new calculation are in red. See text for additional details.}
    \label{fig:hf-comp-all-omp-w-cs}
\end{figure}

The result is a higher cross section at $6.2$ MeV, but overall consistent, when compared to the values extracted from the elastic scattering data alone. The tail of the cross section distribution is significantly larger, leading to a dramatic increase in the overall uncertainty at $6.2$ MeV (approaching a factor of $2$), a clear sign of the tension between our elastics data and the cross section data of Ref.~\cite{oprea-2017}. The \Sran cross section can be seen in Fig.~\ref{fig:hf-comp-all-omp-w-cs} compared to the elastic only data and TALYS optical models. Importantly, no movement was seen in the energy dependent parameters of the linear potential model, indicating these cross section points do not provide any additional information about the energy dependence of the imaginary or real potentials. The shift in the cross section is entirely due to changes in the geometric parameters.

\section{Conclusions}
\label{sec:conclusions}

In this study we have analyzed data from the $^{86}$Sr$(\alpha,\alpha)$ reaction taken at TUNL. Despite the modest precision of the data, we were able to place constraints on the cross section at low energies of around $50 \%$ taking into account experimental and optical model uncertainties using a fully Bayesian analysis. We explored two different methods of extrapolating the potential parameters to lower energies, both of which produced compatible predictions. We also allowed for multiple modes in the real potential (i.e the discrete ambiguity). Our observations run contrary to frequent claims in the literature that various optical model uncertainties make low energy $(\alpha, n)$ cross sections subject to uncontrolled uncertainties, particularly the imaginary potential. The behavior of the \Sran cross section at $E_{\alpha} = 6.2$ MeV seems to be almost entirely determined from the overall imaginary potential strength at radii exceeding $10$ fm. Our data appear to acceptably constrain the potential in this region regardless of the chosen energy dependence, supporting the value and predictive power of elastic scattering data. { These observations only hold for the parametric uncertainties of our \aomp, further work is needed to determine if \textit{model} uncertainties, such as the assumption of the Woods-Saxon potential, play a significant role. Furthermore, as mentioned in Sec.~\ref{sec:stat-model} other Hauser-Feshbach inputs such as level densities can also impact the cross section. It is worthwhile to investigate the extent to which these other model parameters are correlated with \aomp parametric uncertainties. }

The importance of uncertainty quantification on the optical model parameters themselves must be stressed. As shown in Sec.~\ref{sec:cross-sect-pred}, even seemingly compatible sets of parameters can occupy very different regions of the full parameter space and the $(\alpha, n)$ cross section is sensitive to this joint space. Our two models of energy dependence predicted imaginary depths a factor of 4 different from each another, but with almost no change in the predicted low energy cross section. While this is due to compatible overall imaginary strengths at radii larger than $10$ fm, such a deduction is only possible due to the fully correlated samples provided to us from the Bayesian analysis since correlations exist between all \aomp parameters (see Fig.~\ref{fig:hf-comp-corner}). Without properly accounting for these parameter correlations it is impossible to actually assess the differences between optical potentials.

Considering the above points, our first recommendation is that future work focus on quantifying the agreement between reaction cross section and elastics scattering data for the cases where both are available. A Bayesian analysis such as ours should be able to settle whether these data are ultimately consistent given optical model uncertainties. We would also advocate that, when analyzing the elastic scattering data, the statistical model should leave the overall data normalization free and include the possibility for additional, unmeasured scatter in the data points (see Eq.~(\ref{eq:theory-yield}) and Eq.~(\ref{eq:total-uncertainty})). These ingredients ensure that minor deviations from the optical model do not influence the potential parameters too strongly and reduce the required precision for elastic studies making them more feasible with modest experimental setups and for radioactive ion beams in inverse kinematics. Our results indicate that these additional precautionary parameters do not negatively influence inference. In Sec.~\ref{sec:potential-impacts-cross}, we saw that the precise \Sran cross section measurements were able to strongly pull the \aomp parameters away from limits set when considering only the elastic data; it is not clear if adding additional uncertainties, whether they be in overall normalization or statistical scatter, is a good practice for these data. Considering the simplicity of the assumption that $\sigma_{reac} \approx \sigma_{(\alpha, n)}$, it is worth exploring such a procedure in future work.

Our second recommendation concerns the direct measurements of $(\alpha, n)$ cross sections.
While low energy measurements of the cross section can obviously give accurate predictions of the thermonuclear reaction rate, they can be challenging and time consuming to carry out for all of the reactions playing an important role in the weak r-process. Additionally, if these studies provide only a handful of data points of modest precision at higher energies, our investigations indicate little if any constraint will be placed on the \aomp at low energies. The implication has been made that the theoretical cross section from a Hauser-Feshbach calculation can merely be scaled based on these data points or that one of the several \aomp from TALYS which matches the measurement can be adopted. These procedures are not compatible with our findings, since even seemingly discrepant data will push parameters towards values that will explain the cross section at those particular energies, which can cause the uncertainties at low energies to actually \textit{increase}. Furthermore, if no attempt is made to update the \aomp parameter based on such data, then little improvement can be expected in the astrophysical region, especially as uncertainties on experimental cross section approach $50 \%$. It would be preferable for direct measurements to be carried out with the intent of constraining the \aomp parameters, meaning: statistical precision of $< 10 \%$, many data points (say $>3$), and carried out at energies where $\sigma_{reac} \approx \sigma_{(\alpha, n)}$. Ref.~\cite{gyurky_2023} is a good example of this approach.

We made no attempt to calculate the reaction rate for \Sran using our results. Finding an efficient way to do this so that reaction rate libraries that include uncertainties could incorporate experimental results would be advantageous \cite{sallaska_2013}. Finally, construction of a global \aomp with quantified uncertainties similar to the work of Ref.~\cite{Pruitt_2023} is desirable.

\section*{Acknowledgments}

The authors would like to thank the TUNL technical staff for their assistance during the experiment and to Robert Janssens for his careful reading of the manuscript. CM appreciates the fruitful conversations and guidance on optical model calculations provided by Carl Brune. This material is based upon work supported by the U.S. DOE, Office of Science, Office of Nuclear Physics, under grants DE-FG02-97ER41041 (UNC) and DE-FG02-97ER41033 (TUNL). TL acknowledges the support of the US National Science Foundation Grant No. PHY-2150118-01.



\bibliography{paper}{}

\end{document}